\documentclass[10pt,twocolumn]{article} 
\usepackage{simpleConference}
\usepackage{times}
\usepackage{graphicx}
\usepackage{amssymb}
\usepackage{url,hyperref}

\usepackage{subcaption}
\usepackage{orcidlink}
\usepackage{amsmath}
\usepackage{siunitx}

\usepackage{textgreek}
\usepackage{array}
\newcommand{\PreserveBackslash}[1]{\let\temp=\\#1\let\\=\temp}
\newcolumntype{C}[1]{>{\PreserveBackslash\centering}p{#1}}
\newcolumntype{R}[1]{>{\PreserveBackslash\raggedleft}p{#1}}
\newcolumntype{L}[1]{>{\PreserveBackslash\raggedright}p{#1}}

\newcommand{\mytilde}{\raise.17ex\hbox{$\scriptstyle\mathtt{\sim}$}}

\begin{document}

\hypersetup{hidelinks=true}
\widowpenalty10000

\title{Machine Learning Based Compensation for \\ Inconsistencies in Knitted Force Sensors}

\author{Roland Aigner\,\orcidlink{0000-0002-8503-4335} \\ \href{mailto:roland.aigner@fh-hagenberg.at}{roland.aigner@fh-hagenberg.at} 
    \and Andreas Stöckl\,\orcidlink{0000-0003-1646-0514} \\ \href{mailto:andreas.stoeckl@fh-hagenberg.at}{andreas.stoeckl@fh-hagenberg.at}}

\affiliation{Media Interaction Lab, University of Applied Sciences Upper Austria \\ 4232 Hagenberg, Austria}

\maketitle
\thispagestyle{empty}

\begin{abstract}
Knitted sensors frequently suffer from inconsistencies due to innate effects such as offset, relaxation, and drift. These properties, in combination, make it challenging to reliably map from sensor data to physical actuation. In this paper, we demonstrate a method for counteracting this by applying processing using a minimal artificial neural network (ANN) in combination with straightforward pre-processing. We apply a number of exponential smoothing filters on a re-sampled sensor signal, to produce features that preserve different levels of historical sensor data and, in combination, represent an adequate state of previous sensor actuation. By training a three-layer ANN with a total of 8 neurons, we manage to significantly improve the mapping between sensor reading and actuation force. Our findings also show that our technique translates to sensors of reasonably different composition in terms of material and structure, and it can furthermore be applied to related physical features such as strain. \end{abstract}

\section{Introduction}
\label{sec:introduction}


Textile based sensors are of high interest in research and industry due to numerous beneficial properties, such as lightness, breathability, and potential stretchability. In particular, knits are inherently elastic textiles, due to their geometric composition of courses of interlocking loops, as opposed to weaves, where yarn is travelling straight. This elasticity makes them ideal for sensing stress or strain \cite{
Scilingo2003,Liang2022,Alam2022,Yang2023} that generally perform according to Holm's theory \cite{Holm1967}, which states that contact resistance $R$  depends on material resistivity $\rho$, material hardness $H$, contact point count $n$, and pressure $P$, with
    $$R=\frac{\rho}{2}\sqrt{\frac{\pi H}{n P}} \,.$$
Consequently, the overall sensor resistance drops when pressure at the loop intermeshing points' contacts is increased, e.g., by straining or pressing. However, depending on the yarn material properties and/or structural composition of a knitted structure, knitted sensors usually suffer from considerable inconsistencies that have to be addressed \cite{Zhang2006,Grassi2017,Bozali2021,Alam2022}, such as settling effects, offset, overshooting, hysteresis, as well as long- and short-term sensor drift, some of which we speculate are due to slight structural re-arrangements of the yarn within the fabric. Inherently, the raw measurement signal tends to seriously deviate from the desired output, i.e., the applied force. This is undesirable in several use cases since it complicates downstream analysis of raw sensor data. Unfortunately, since these effects are unlike common noise, traditional methods such as frequency-domain- or Kalman-filters are insufficient to get rid of those.

\begin{figure*}[t]
    \centering
    \begin{subfigure}{1\textwidth}
      \centering
      \includegraphics[width=1\textwidth]{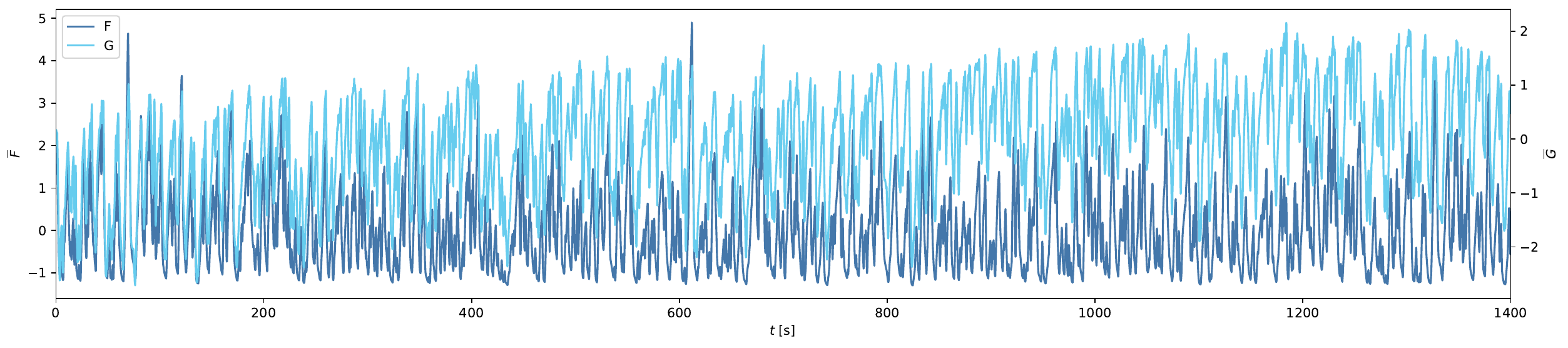}\
    \end{subfigure}
    \begin{subfigure}{.5\textwidth}
      \centering
      \includegraphics[width=1\textwidth]{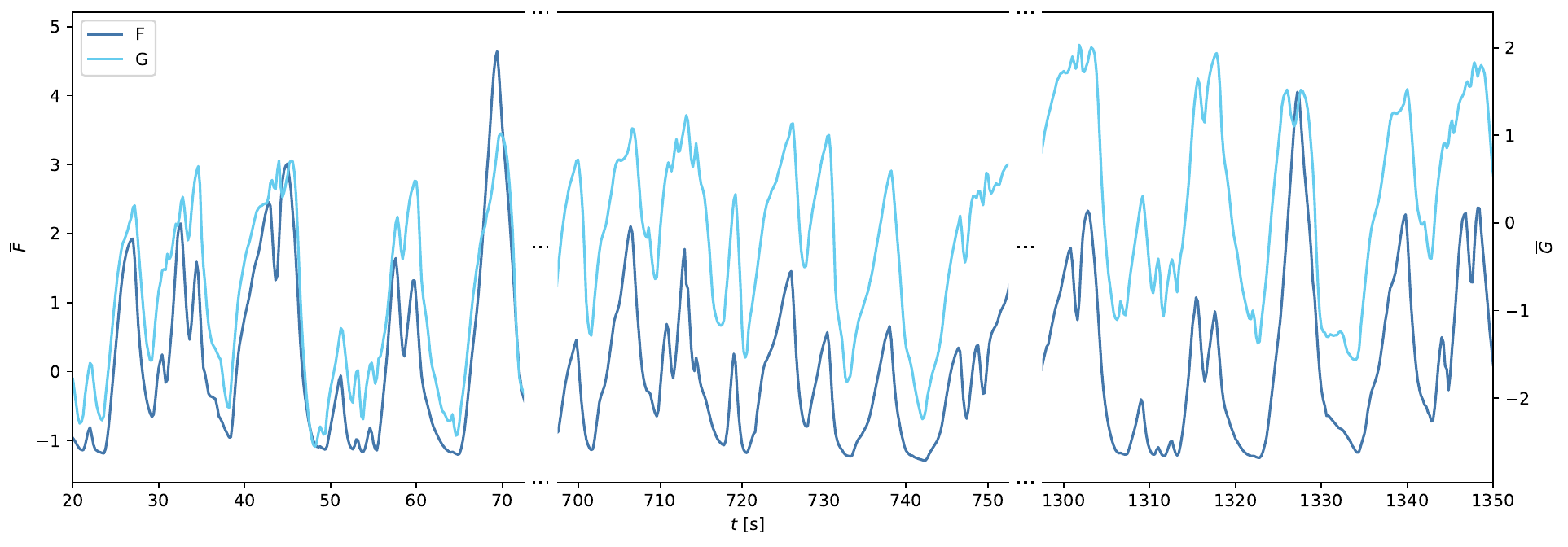}
    \end{subfigure}%
    \begin{subfigure}{.5\textwidth}
      \centering
      \includegraphics[width=1\textwidth]{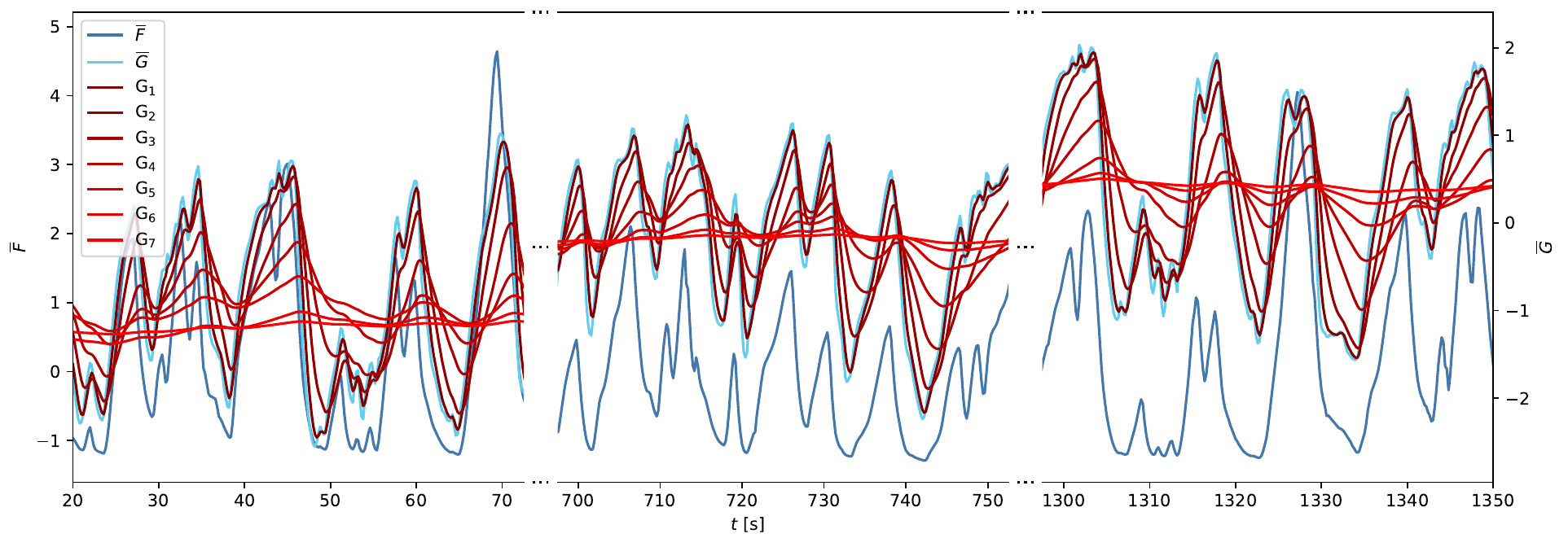}
    \end{subfigure}
    \caption{Timeline of one of our recordings using the PES sensor (top): Overlaying normalized trends of ground truth $\overline{F}$ and sensor readings $\overline{G}$ clearly shows an upwards drift of the sensor readings. Zooming into three segments of the timeline (bottom left), reveals inconsistencies and even contradicting effects, such as overshooting vs. underestimation in the first segment. Our method utilizes a neural network that is trained using several smoothed sensor signals as features (bottom right). We combine a set of those with different smoothing factors to incorporate different degrees of historic information.}
    \label{fig:timeline}
\end{figure*}

Figure \ref{fig:timeline} (top) shows an example timeline plot of a recording of applied force and resulting sensor reading. Long-term drift is most eminent, however short-term inconsistencies are also apparent when zooming in (cf. Figure \ref{fig:timeline} bottom left, for a magnification of three timeline snippets), which illustrate that a basic mapping, e.g., by multi-point calibration is impractical and not promising. Due to the nature of a knitted fabric, effects like latency, hysteresis, drift, offset, overshooting, etc. (and the interaction thereof) are innate. Furthermore, as their extent depends also on the chosen knitting structure, which makes their use challenging for scenarios where reliability is required. 

Our main hypothesis for this work is that inconsistencies that are reflected in the sampling data are in fact ultimately deterministic, as their cause is in the knits' physical and geometrical composition, however too complex to analyze or model manually from empirical observations. Hence, our approach is to utilize an artificial neural network (ANN) to learn and model those factors instead. A further objective is to keep the computational complexity low by reducing the number of required features and by using small-scale networks, so our method is viable for low-end embedded devices with highly limited computational capabilities.

In this paper, we present an easy-to-implement method for mitigating this error, which is based on a small-scale multilayer perceptron neural network (MLP NN) and comes with little modeling/training effort and low computational cost. NNs are in general frequently used for modeling complex non-linear systems, as it has been shown that even small networks are able to well approximate any continuous function of $n$ real variables \cite{
Park1991}. Moreover, MLP networks are relatively easily trained using Backpropagation (BP) algorithms \cite{Russell2010
}.

The main contributions of this paper are as follows: 
\begin{itemize}
    \item a method for rectifying inherent inconsistencies in raw sensor data that result from the structural nature of knitted sensors.
    \item description of a related feature-set that we used as an input vector for an ANN that model several levels of historic sensor data.
    \item mitigation of short-term errors, was well as removal of long-term sensor drift, that go beyond sensor hysteresis.
    \item an exemplary processing pipeline including a NN of minimal complexity that is easily trained and computationally undemanding during operation.
    \item a demonstration of the method's transferability to sensor knits of different structure and to different objectives (e.g., mapping sensor data to stain instead of stress).
\end{itemize}



While there are numerous works that used Machine Learning techniques for classification, e.g., for detecting hand gestures \cite{Lee2021}, sitting postures \cite{Jiang2022}, or exercise activities \cite{Wicaksono2022}, others have used neural networks in scenarios similar to ours, however mostly for compensating hysteretic behavior. Dang et al. \cite{Dang2005} used Radial Basis Function (RBF) NNs to model Preisach hysteresis of piezoceramic actuators. Similarly, Lien et al. \cite{Lien2010} used hysteretic recurrent NNs in the context of piezoelectric actuators, modeling hysteresis in the neurons' activation function. Tong et al. \cite{Tong2005} used Backlash-Based Hysteresis Simulation Models to test a NN that approximates hysteretic non-linearities. Wu et al. \cite{Wu2009} implemented a dynamic NN structure based on the Hammerstein model for dynamic error compensation of infrared thermometer sensors. More recently, Weiss et al. \cite{Weiss2018} applied Kalman filters for preprocessing chemical sensors' data for downstream machine learning. Jondhale et al. \cite{Jondhale2019} combined Kalman filters with General Regression NNs for 2D-position tracking from RSSI signals. In the field of textile-based sensing, Atitallah et al. \cite{Atitallah2021} compared filtering methods such as moving average, moving median, Savitzky-Golay, and Gaussian for processing data from a sensor glove that incorporated CNT-based sensors. Vu et al. \cite{Vu2020} implemented an adaptive fuzzy-NN for capacitive pressure sensors, which were based on spacer-knit structures. Finally, Liu et al. \cite{Liu2021} utilized RBF NNs to compensate for hysteresis disturbance in non-affine, nonlinear systems, however the test set consists of limited, generated data that is furthermore repetitive.

While all those works are related, objective, material, and use cases differ from ours and hence the techniques are hardly transferable. To our knowledge, there is so far no related work applying NNs for data rectification of knitted piezoresistive stress/strain sensors targeted at random, real-world actuation.

\section{Sensor Implementation}
\label{sec:sensor}



\begin{figure*}
    \centering
    \begin{subfigure}[t]{.5\textwidth}
      \centering
      \includegraphics[width=1\linewidth]{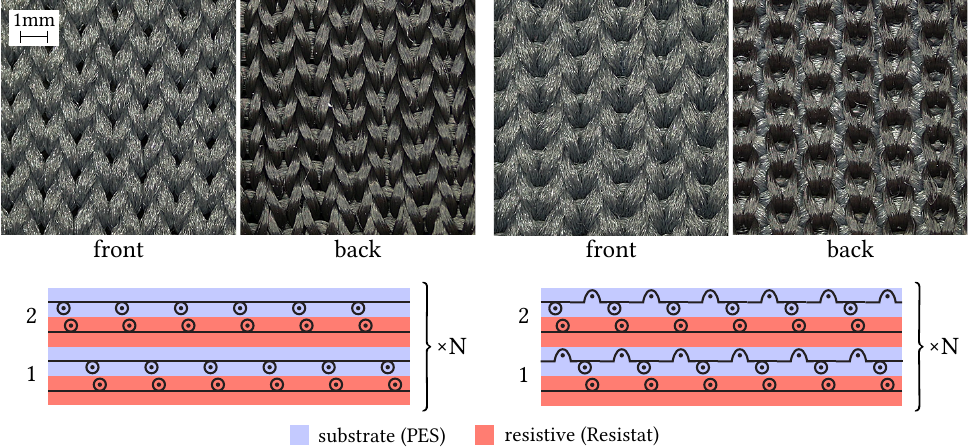} 
      \caption{}
      \label{fig:sensors:sub1}
    \end{subfigure}
    \begin{subfigure}[t]{.22\textwidth}
      \centering
      \includegraphics[width=1\linewidth]{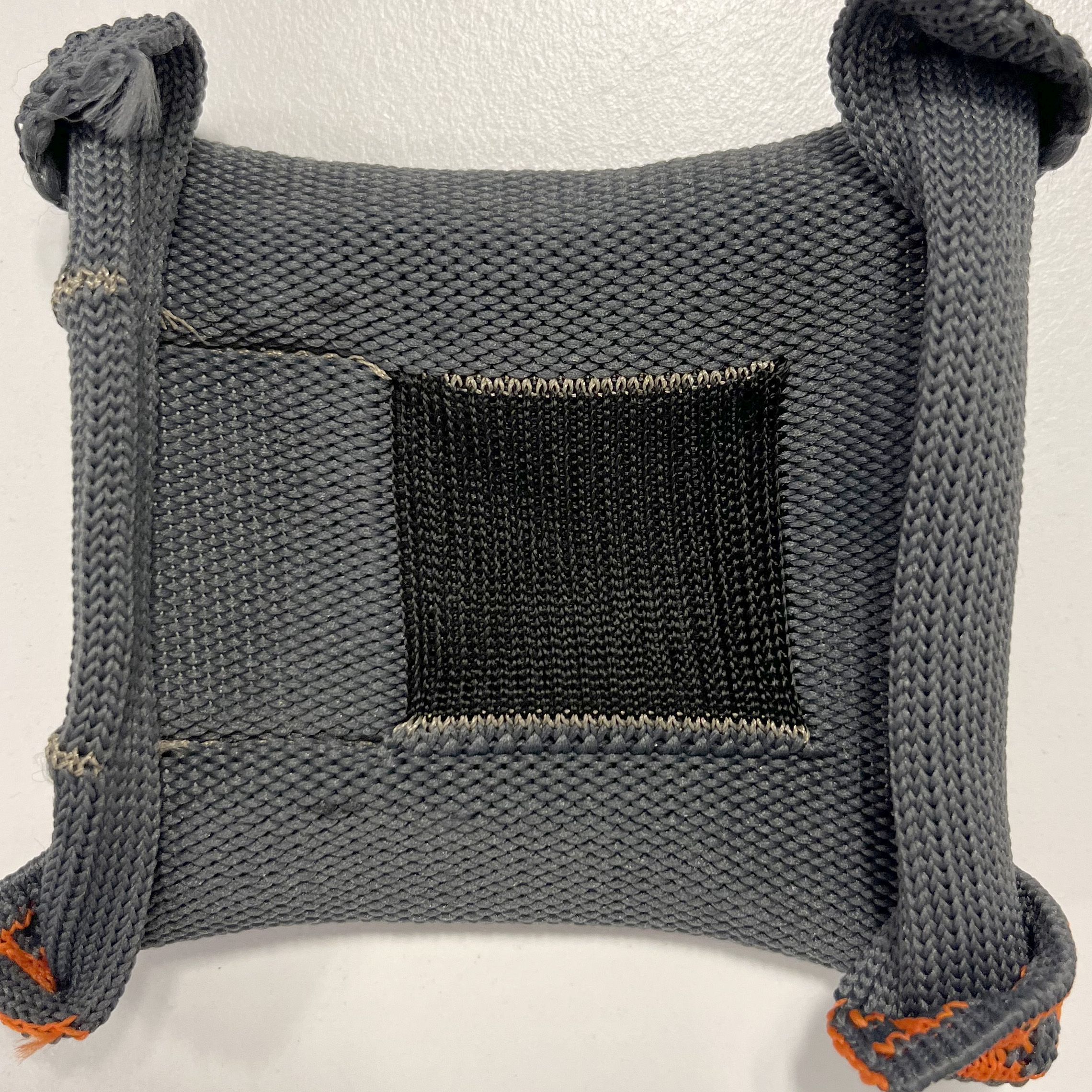}
      \caption{}
      \label{fig:sensors:sub2}
    \end{subfigure}
    \begin{subfigure}[t]{.22\textwidth}
      \centering
      \includegraphics[width=1\linewidth]{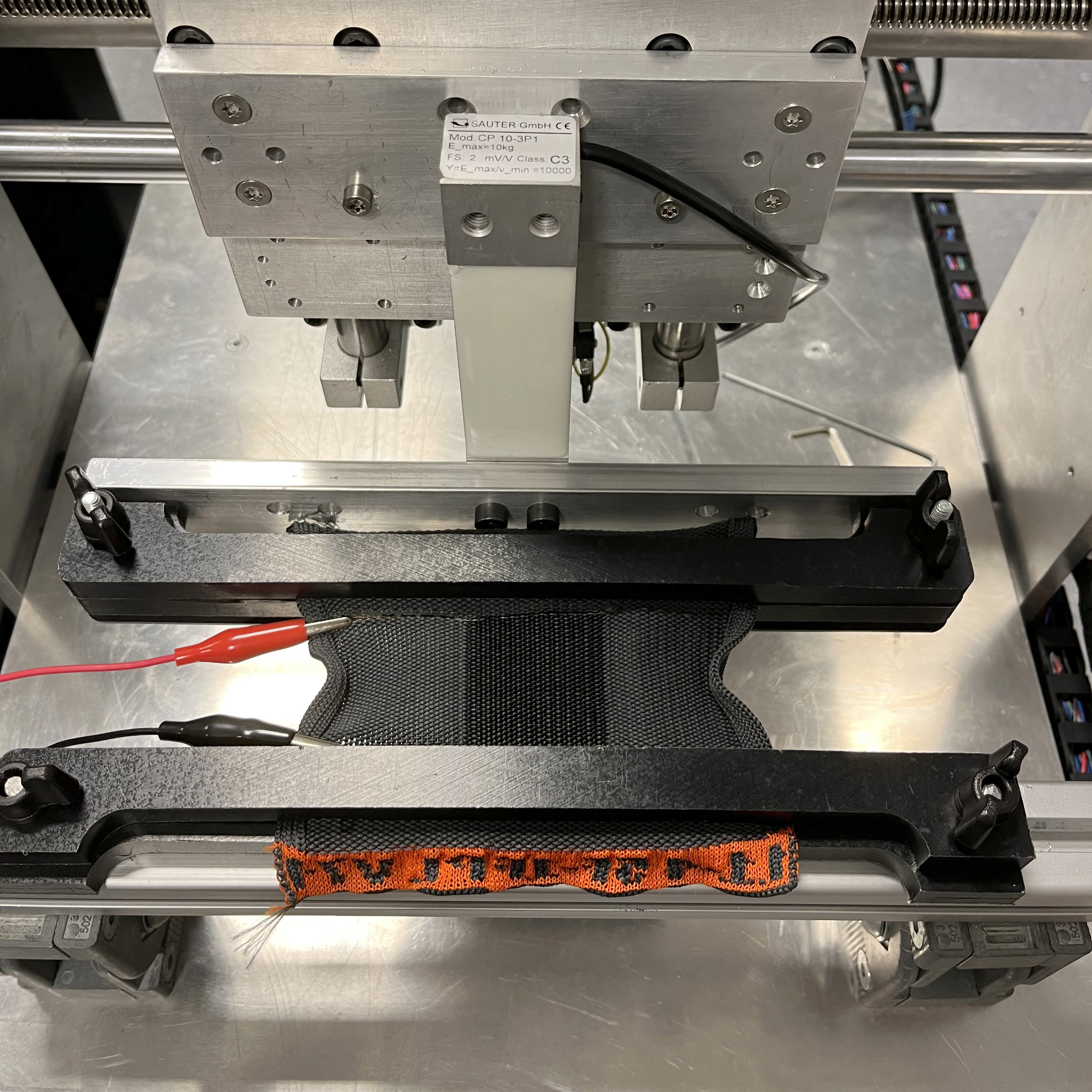}
      \caption{}
      \label{fig:sensors:sub3}
    \end{subfigure}
    \caption{Closeups and knitting patterns of the Twill based knitting structures (a): for the PES- (tubular structure, left), and Lycra (connected, i.e., PES tucked to back-face Resistat, right). PES sensor patch (b), with conductive yarn traces connecting the resistive area (black) on both upper and lower ends. We evaluated our sensors using a custom-built tensile tester which is equipped with a force cell (c).
    }
    \label{fig:sensors}
\end{figure*}

Our sensors utilize a widespread knitting pattern that is often, however inconsistently, called "Twill" in the textile industry, due to its structural similarity with a Twill Weave. It consists of courses with alternating knit and float stitches, shifted by one needle every other row (cf. Figure \ref{fig:sensors:sub1})
. The high number of floats results in exceptional stability along course-direction and high elasticity along wale-direction, when compared to more straightforward knits, such as Plain or Double Jersey \cite{Spencer2001}. In this regard it exhibits a characteristic similar to a Cardigan, however with orthogonal anisotropic behavior, which can be an advantage in certain use case scenario, where omni-directional elasticity is not desired. 


Since a Twill can be knitted on a single needle bed, the opposite bed is available for additional structures. We decided to apply our knitted sensors on a substrate carrier structure, by knitting a Twill consisting of PES on the front bed and another Twill using resistive yarn on the back bed, to have the sensor part completely covered by the PES on one side for protecting it from abrasion. Furthermore, this gave us better control in balancing for a uniform stability across the whole fabric, when compared to integrating a sensor patch as an Intarsia \cite{Lee2021}. Note that this requires the two faces to be connected to not fall apart; we did this by tucking the resistive yarn to the PES at the outer wales of the sensor area produces a tubular structure (cf. Figure \ref{fig:sensors:sub1}, left).


From a previous study \cite{Aigner2023}, we learned that adding Lycra to the substrate can significantly improve elastic recoil, minimizing hysteresis. We therefore fabricated two variations: one with pure PES-substrate, one with additional Lycra. Since the resistive face is not complemented with Lycra, we tightly connected front and back faces for these sensor patches, by tucking the substrate to the sensor knit for every loop (cf. Figure \ref{fig:sensors:sub1}, right), to prevented interference from resistive face's lagging behind the elastic substrate. We decided to include this additional version in our evaluation, to estimate how well our method translates to sensors of different design.





For the substrate carrier, we used a den\,150 PES from TWD Fibres GmbH, and for the Lycra thread we used a \,140 Lycra core covered with PES den\,150/20 from Jörg Lederer GmbH. For the resistive sensing areas, we used Polyester-based, Carbon-sheathed Resistat P6204\footnote{\url{https://shakespeare-pf.com/product/polyester/}} from Shakespeare\textsuperscript{\textregistered} with den\,100/24 and \mytilde 10\,M$\Omega$/m. For the connector traces, we used silver-coated PA-yarn Madeira HC40\footnote{\url{https://www.shieldex.de/products/madeira-hc-40/}}, with den\,260 and \textless 300\,$\Omega$/m. All our patches were knitted on a flat-bed knitting machine of type ADF 530-32 KI W Multi Gauge from KARL MAYER STOLL, at gauge E\,7.2. For more details regarding materials and fabrication of our sensors, we refer to \cite{Aigner2023}.


\section{Data Acquisition}


For collecting training and test data, we used a custom-made tensile tester, which we built from a CNC milling machine (cf. Figure \ref{fig:sensors:sub3}). The clamps for attaching the patches at both ends featured needles at 2\,cm distance to secure the textile against slipping. The moving actuator was equipped with a single-point load-cell with nominal load of 10\,kg (Sauter CP 10-3P1\footnote{\url{https://www.kern-sohn.com/shop/en/products/measuring-technology-components/CP-10-3P1/}}) and was sampled at \mytilde 40\,Hz by an ADS\ 1231 24-bit Delta-Sigma ADC\footnote{\url{https://www.ti.com/product/ADS1231}}. Since sensor resistance readings were slightly noisier, we supersampled with 128\,Hz via a simple voltage divider with a 606\,k$\Omega$ reference resistor. We used an Adafruit ADS1115 16-bit ADC\footnote{\url{https://www.adafruit.com/product/1085}} for sampling, buffered readings and averaged values in windows of \mytilde 25\,ms, again resulting in a rate of \mytilde 40\,Hz for our final samples. Measurements for force and resistance, as well as actuator displacement and timestamps were captured into CSV files by the MCU firmware. A single ESP32 on an Adafruit HUZZAH32 Feather board\footnote{\url{https://www.adafruit.com/product/3405}} was used for sampling and recording to SD card. 

\begin{figure}
    \centering
    \includegraphics[width=1\linewidth]{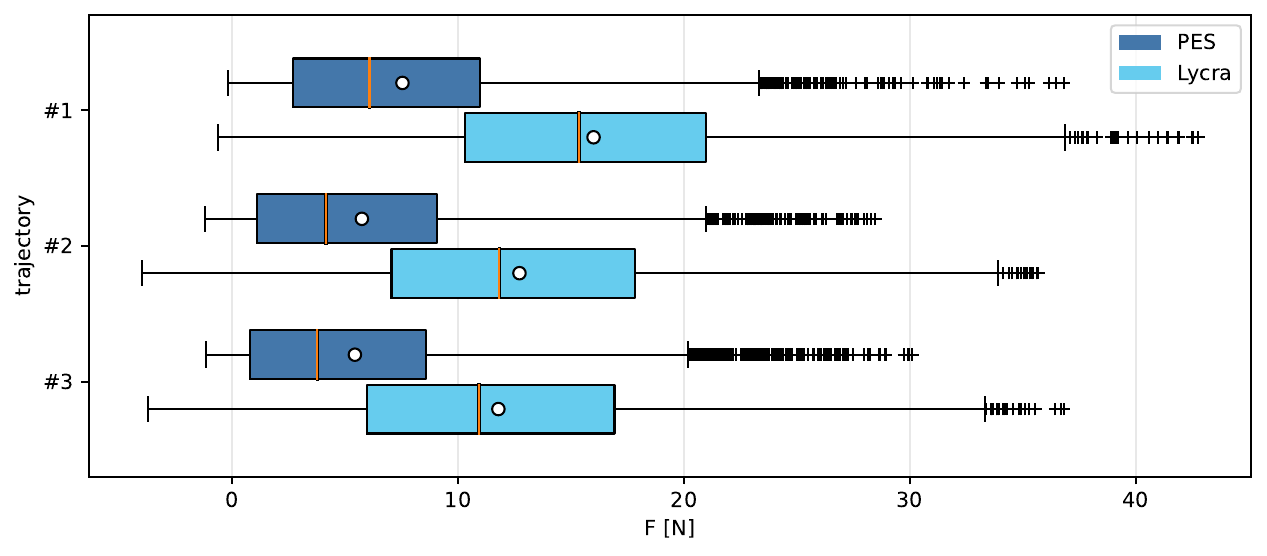}
    \caption{Due to different knitting structures and material compositions, the two sensors operate in different force ranges when actuated using identical trajectories.}
    \label{fig:trajectory}
\end{figure}

The tensile tester was controlled by Art-Soft Mach4 CNC Control Software (v4.2.0), running on a Windows\ 10 PC. To simulate pseudo-natural motion for reasonably representative actuation of the sensors and collecting corresponding sensor data, we generated G-code trajectories based on Perlin Noise \cite{Perlin1985,Perlin2002}, to control the actuator along the x-axis. Our objective was to generate non-repetitive trajectories, so we could guarantee our resulting model would not merely learn to repeat a specific pattern and instead be applicable for random actuation. We created three of those trajectories, to get one data set for training our ML models and two for testing their performance. Based on pre-evaluation of the sensors  \cite{Aigner2023}, we chose the amplitudes of our trajectories to move in an approximate range of 0\% to 30\% extension. Maximum, mean, and SD of velocities and accelerations can be found in Table \ref{tab:trajectory}. We collected data for both of our sensors, i.e., PES-only ("PES") and Lycra-enhanced ("Lycra") variants. Note that this resulted in different force ranges, due to the difference in firmness (cf. Figures \ref{fig:trajectory}, bottom). Our objective was to test our method on sensors of slightly different structure and material, to get an estimation of its portability between sensor designs. The total duration of our actuation process and the time-span we recorded was \mytilde 23 minutes each.

\begin{table}
    \centering
    \caption{Statistics of our three generated actuator trajectories. Velocities were around 1\,mm/s, with a maximum of 4.930\,mm/s.}
    \label{tab:trajectory}
    \begin{tabular}{r|c|c|c|c|c|c}
                   & \multicolumn{3}{c|}{v [mm/s]} & \multicolumn{3}{c}{a [mm/s²]} \\
        traj. & max & mean & SD & max & mean & SD \\
    \hline
        \#1 & 4.142 & 0.943 & 0.593 & 19.05 & 1.397 & 1.303 \\
        \#2 & 4.930 & 0.979 & 0.617 & 18.61 & 1.497 & 1.343 \\
        \#3 & 4.039 & 1.030 & 0.644 & 19.25 & 1.546 & 1.414 \\
    \end{tabular}
\end{table}

\section{Data Processing}
\label{sec:processing}

For reasons of simplicity, our presented method was not designed to take timing into account yet, hence it requires uniform sampling periods. However, we noticed the sampling frequency was not perfectly constant (\textmu=41.5\,Hz, \textsigma=14.2), as a consequence of our multi-component setup. We further noticed the targeted frame rate of \mytilde 40\,Hz was higher than what our method required, since results did not significant change after downsampling to 20\,Hz. Resampling with even lower rates (e.g., 10\,Hz) seemed to yield worse results, though, we therefore re-sampled our recorded force and resistance data to 20\,Hz using linear interpolation in between the sample points. 

The main objective of our work is to infer force-data from the raw measurements that were taken from the sensor, meaning both trends should be as identical as possible, which is not the case with raw measurement data (cf. Figure \ref{fig:timeline}). Hence, we argue the coefficient of determination $r^2$ is a reasonable metric for quantifying this property and hence to judge about the performance of our approach\footnote{Note that we use the less-common lower-case notation $r^2$ to avoid confusion with the sensors' electrical resistance $R$.}. However, since force $F$ and sensor resistance $R$ are inversely proportional, we utilize the sensor conductivity $G=1/R$ instead of the resistance. Furthermore, since both $F$ and $G$ cover largely different ranges, we normalized both to identical ranges to $\overline{F}$ and $\overline{G}$, by removing mean and scaling to unit variance using the \textit{StandardScaler} from the \textit{scikit learn} Python package\footnote{\url{https://scikit-learn.org/stable/modules/generated/sklearn.preprocessing.StandardScaler.html}}. This pre-processing step is also beneficial (and in fact recommended \cite{SKLearnStdScaler}) for better performance of machine learning estimators later on. Hence, we calculate the coefficient of determination with 

$$
    r^2(X,Y) = 1 - \frac{\sum_i \left( x_i - y_i \right)^2}{\sum_i \left( x_i - \texttt{mean}(Y) \right)^2} \,
$$
substituting $X$ with normalized force measurement $\overline{F}$, and $Y$ with pre-processed and normalized conductivity measurement $\overline{G}$. To quantify the performance of our machine learning model, we then use the prediction $p$ for $Y$ instead.

After pre-processing our raw data, we calculated our initial $r^2$ scores as baseline values with 0.423, 0.471, and 0.526 for the recordings with our PES version and as 0.703, 0.667, and 0.734 for those with the Lycra version (cf. Table \ref{tab:prediction}). It is already eminent that the performance of our Lycra patch is superior, which is in line with previous findings  \cite{Aigner2023}. However, we first focus on the most basic implementation, without additional Lycra, which we expect to benefit most from improvement by computational means.

\begin{table}
    \centering
    \caption{Smoothing base value \textit{a} and factor count \textit{N} were varied for determining factors $\alpha_i$, e.g., a = 10 and N = 3 result in A = (10\textsuperscript{-1}, 10\textsuperscript{-2}, 10\textsuperscript{-3}). The values for our initial factors that we used as baseline were adjusted manually and therefore not calculated from $a$ and $N$.}
    \begin{tabular}{cc|ll}
        $a$ & $N$ & $A = (\alpha_1 \ldots \alpha_N)$ & \\
        \hline \hline
        - & - & (0.5, 0.1, 0.025, 0.0025) & baseline \\
        \hline
        2.5 & 4  & (0.4, 0.16, 0.064, 0.0256) & $\alpha_i = 1/a^i$ \\
        2.5 & 7  & (0.4, 0.16, \textellipsis 0.0016384) & \\
        2.5 & 10 & (0.4, 0.16, \textellipsis 0.000104858) & \\
        5   & 4  & (0.2, 0.04, 0.008, 0.0016) & \\
        5   & 7  & (0.2, 0.04, \textellipsis 0.0000128) & \\
        10  & 3  & (0.1, 0.01, 0.001) & \\
        10  & 4  & (0.1, 0.01, 0.001, 0.0001) &
    \end{tabular}
    \label{tab:smoothingfactors}
\end{table}

As mentioned, our hypothesis is that an ANN is able to model the knit's state and infer the actuation from an (seemingly randomly) inconsistent sensor reading. For example, it seems reasonable that sensor offset and settling speed is affected by recent elongation. From that follows, that there needs to be historic data available for the current prediction; this can either be implemented by a feedback mechanism, or otherwise by choosing input features that include temporal information. For sake of simplicity, we decided for the latter. Using a number of $n$ previous samples within a certain time-frame as features is not promising since this would result in a very high number of features and would therefore rapidly increase complexity of computation and network topology, increasing the risk of overfitting. Furthermore, this approach would highly depend on the sample-rate. Instead, we decided for providing historic data in the form of several smoothed signals, with varying degrees of responsiveness. We utilized exponential smoothing \cite{Gardner1985} with 
    $$y(t) = \alpha x_t + (1-\alpha) y(t-1)\,,$$
where $\alpha$ is a smoothing factor in range [0 1]; i.e., the lower the value for $\alpha$, the higher the drag. We noticed initialization with $y(0) = x_0$ introduced too much bias for the signals with high drag, therefore we initialized with the mean of samples values within a window of M samples, $y(0) = \texttt{mean}(x_0 \ldots x_{M-1})$. For calculating an adequate window-size that is depending on drag and sample rate $f$, we empirically found $M = \lceil 1 / f \alpha \rceil$ delivers reasonable initialization values. By filtering the sensor conductivity $\overline{G}$ with a set of smoothing factors $\alpha_i$, we gain a set of $N$ filtered sensor signals $G_i$, reflecting different degrees of temporal data (cf. Figure \ref{fig:timeline}, bottom right), which represent the elements of our feature vector. Note that our exponential smoothing implementation does not take timing into account, which is a further reason we re-sampled our data to a constant rate of 20\,Hz.

\begin{table}
    \centering
    \caption{
    Number of hidden layers' (HL) neurons were determined by building the products of \textit{base sizes} with \textit{topology} vectors and flooring the results, e.g., $\lfloor$6 $\times$ (\textonehalf, 1, \textonequarter)$\rfloor$ would give three layers with 3, 6, and 1 neurons, respectively. After removing all identical permutations as well as those containing sizes \textless 2, 114 unique variations remained.}
    \begin{tabular}{r|ll}
        parameter & variations & \\
        \hline \hline
        HL base sizes  & \multicolumn{2}{l}{2, 3, 4, 6, 8, 12, 16, 32} \\
        \hline
        topologies    & \multicolumn{2}{p{6cm}}{
            (   1,   1 ), (   1, \textonehalf\; ), (   1, \textonequarter\; ), \newline
            (   1,   1,   1 ), (   1,   1, \textonehalf\; ), (   1, \textonehalf\;, \textonehalf\; ), (   1, \textonehalf\;, \textonequarter\; ), \newline
            ( \textonehalf\;,   1,   1 ), ( \textonehalf\;,   1, \textonehalf\; ), ( \textonehalf\;,   1, \textonequarter\; ), \newline
            (   1,   1,   1,   1 ), (   1,   1,   1, \textonehalf\; ), (   1,   1, \textonehalf\;, \textonehalf\; ), \newline 
            (   1, \textonehalf\;, \textonehalf\;, \textonehalf\; ), (   1, \textonehalf\;, \textonehalf\;, \textonequarter\; ), (   1, \textonehalf\;, \textonequarter\;, \textonequarter\; ), \newline
            (   1, \textonequarter\;, \textonequarter\;, \textonequarter\; ), ( \textonehalf\;,   1,   1,   1 ), ( \textonehalf\;,   1,   1, \textonehalf\; ), \newline 
            ( \textonehalf\;,   1, \textonehalf\;, \textonehalf\; ), (   1, \textonehalf\;, \textonehalf\;, \textonehalf\; )
           }
    \end{tabular}
    \label{tab:parameters}
\end{table}

Multi-layer Perceptrons (MLP) trained by back-propagation (BP) algorithms are commonly used for function approximation, we therefore used the \textit{MLPRegressor}\footnote{\url{https://scikit-learn.org/stable/modules/generated/sklearn.neural_network.MLPRegressor.html}} of \textit{scikit learn}, with \textit{relu} activation function and maximum iterations of 10.000. Note that MLPs are particularly sensitive to feature scaling \cite{SKLearnNN}, which makes our previously described pre-processing mandatory.

We started with experimental $\alpha$ values of 0.5, 0.1, 0.025, and 0.0025, which gained promising results, therefore we kept this set as a baseline. From there, we tried different sets of smoothing factor vectors $A = (\alpha_1 \ldots \alpha_N)$, with $\alpha_i = 1 / a^i$, modifying $a$ and $N$ to get several sets of different sizes and granularity. Since we found that data smoothed with $\alpha$ values below 10\textsuperscript{-4} held too little information, we mostly refrained from going beyond those. Apart from feature vectors, we varied the ANN's hidden layer \textit{sizes} (i.e., neuron counts) and \textit{topologies}. We did not go beyond neuron counts of 32 and beyond 4 hidden layers, since we started to notice frequent overfitting at these values. We then empirically tested all permutations\footnote{after removing duplicates and those including layers with less than 2 neurons, both due to flooring the products, 114 network permutations remained. Testing those with all 8 variations of A, this resulted in 912 candidates for our pipeline.} of those parameters to find the optimal configuration (cf. Table \ref{tab:parameters} for a complete listing).

\begin{table}
    \centering
    \caption{$r^2$ of our initial (pre-processed) and predicted data. We included results of our two test sets (A, B), as well as the training sets (T) for sake of completeness. PES show highest gain from our approach; however, the Lycra patches ultimately yield higher scores.}
    \begin{tabular}{L{0.1\columnwidth}|C{0.15\columnwidth}|C{0.15\columnwidth}|C{0.15\columnwidth}}
         & r²(F,G) & r²(F,p) & gain \\
        \hline \hline
        PES\textsubscript{t}   & 0.423 & (0.781) & (85\%) \\ 
        PES\textsubscript{A}   & 0.471 & 0.791 & 68\% \\ 
        PES\textsubscript{B}   & 0.526 & 0.767 & 46\% \\ 
        \hline
        Lycra\textsubscript{t} & 0.703 & (0.841) & (20\%) \\ 
        Lycra\textsubscript{A} & 0.667 & 0.830 & 24\% \\ 
        Lycra\textsubscript{B} & 0.734 & 0.828 & 13\% 
    \end{tabular}
    \label{tab:prediction}
\end{table}

\section{Evaluation and Discussion}
\label{sec:evaluation}

\subsection{Results}



Our tests showed that a low number of hidden layers worked well in most cases and increasing them did not considerably improve scores. Occasionally, adding a fourth hidden layer did even degrade results, but this seems also to be subject to network topology. Overall, hidden layer \textit{base sizes} of at least 4 were required, otherwise the resulting models would be unusable. In terms of features, we saw that smoothing factors resulting from $a$ = 10 (with both $N$ = 3 and $N$ = 4) did not perform well. We speculate this is because the smoothing factors are too far apart and the resulting low number of $G_i$ features include too little historical information. Overall, our variations with $a$ = 5 performed better, however the best scores were achieved with $a$ = 2.5 and $N$ \textless 10. A spreadsheet including all the scores of our network variations can be found in the supplementary material.

\begin{figure}[!b]
    \centering
    \includegraphics[width=1\columnwidth]{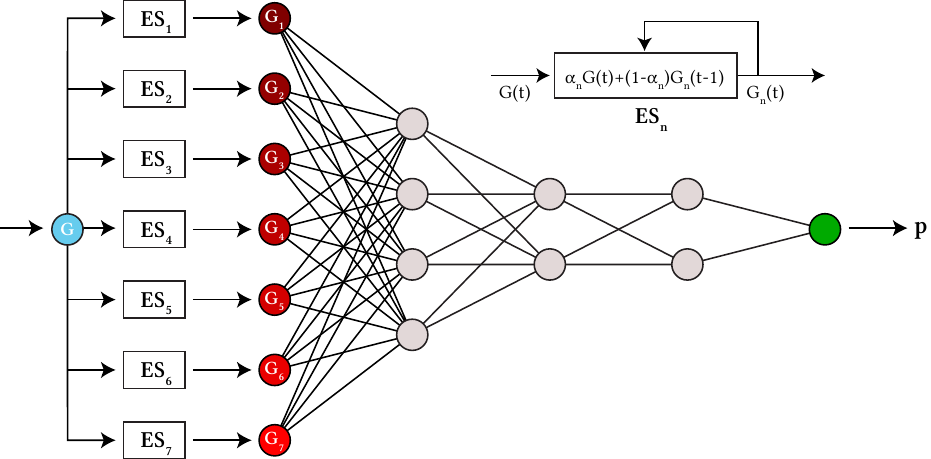}
    \caption{We feed pre-processed (re-sampled and normalized) data $\overline{G}$ into a number of exponential smoothing filters using different smoothing factors and use the results as feature vector for our neural network. In our tests, the combination of a = 2.5 and N = 7 with hidden layer sizes (4, 2, 2) resulted in the best $r^2$ score.}
    \label{fig:ann}
\end{figure}

\begin{figure*}[!t]
    \centering
    \begin{subfigure}{\textwidth}
        \centering
        \includegraphics[width=1\textwidth]{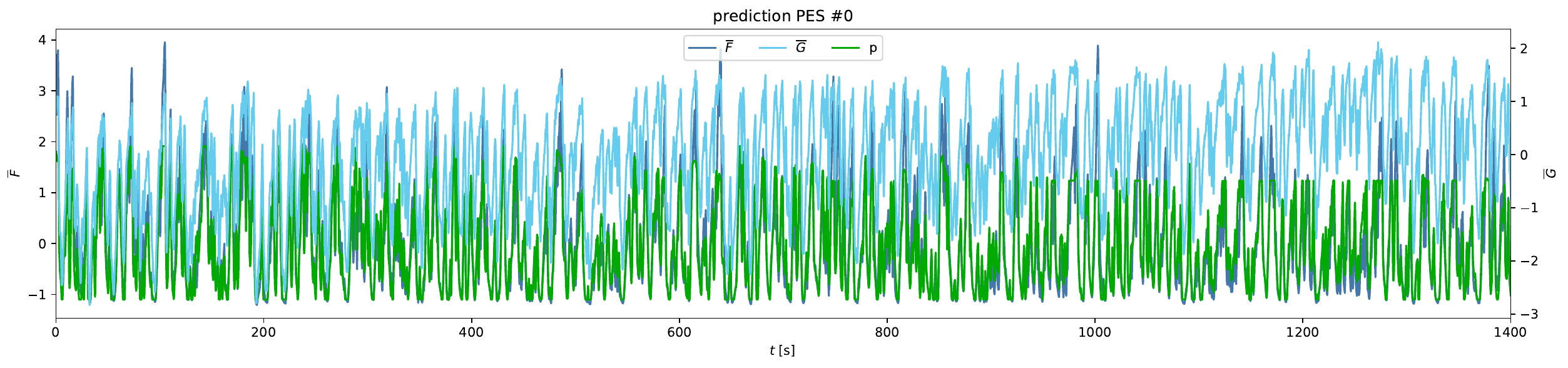}\
    \end{subfigure}
    \begin{subfigure}{.5\textwidth}
        \centering
        \includegraphics[width=1\columnwidth]{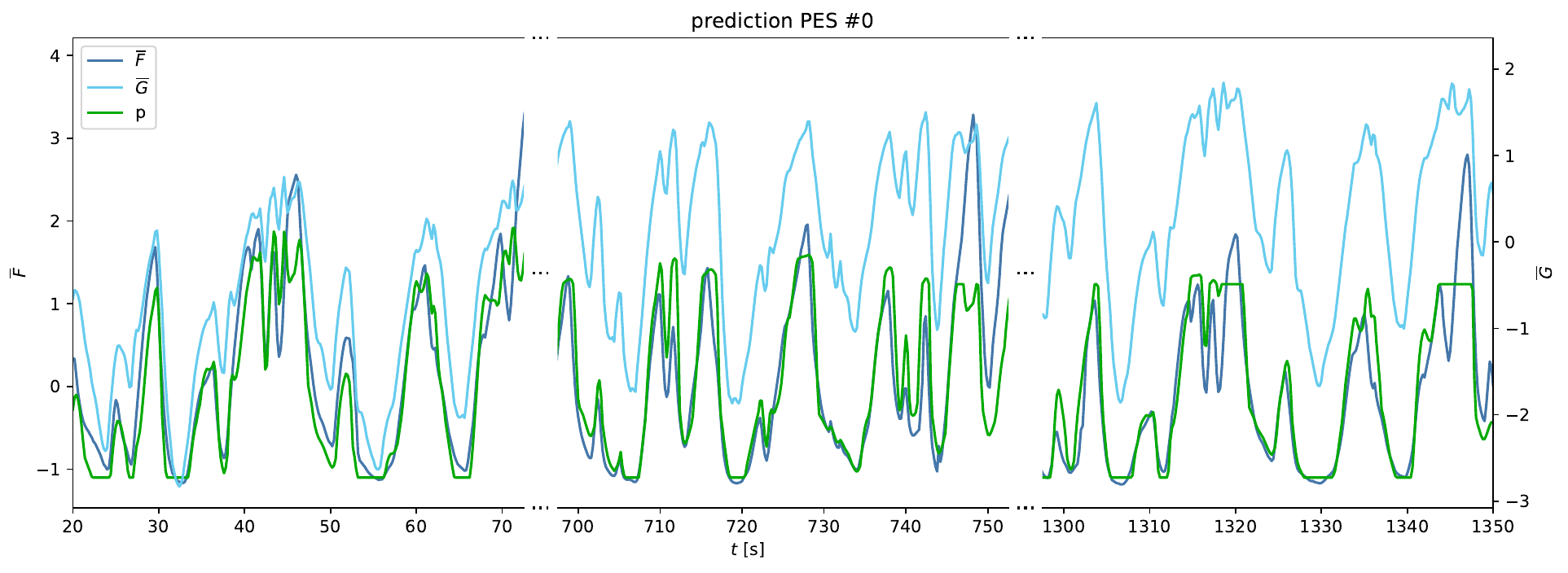}\
        \includegraphics[width=1\columnwidth]{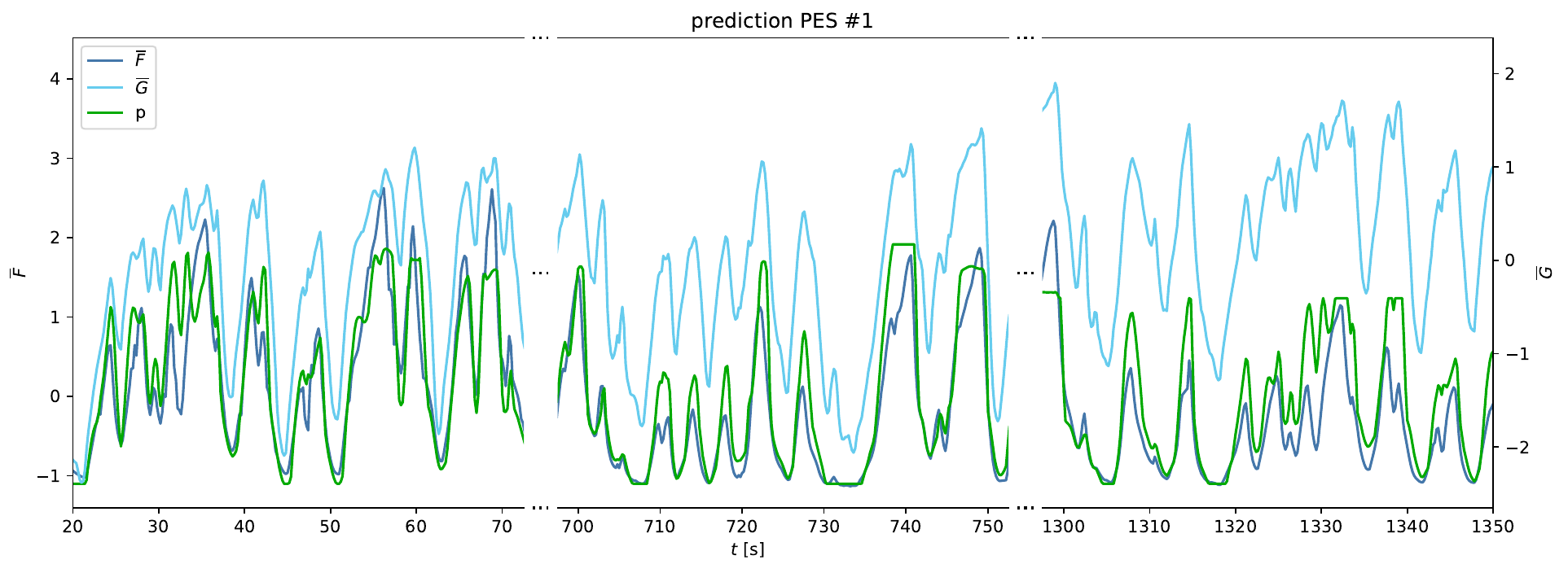}%
    \end{subfigure}%
    \begin{subfigure}{.5\textwidth}
        \centering
        \includegraphics[width=1\columnwidth]{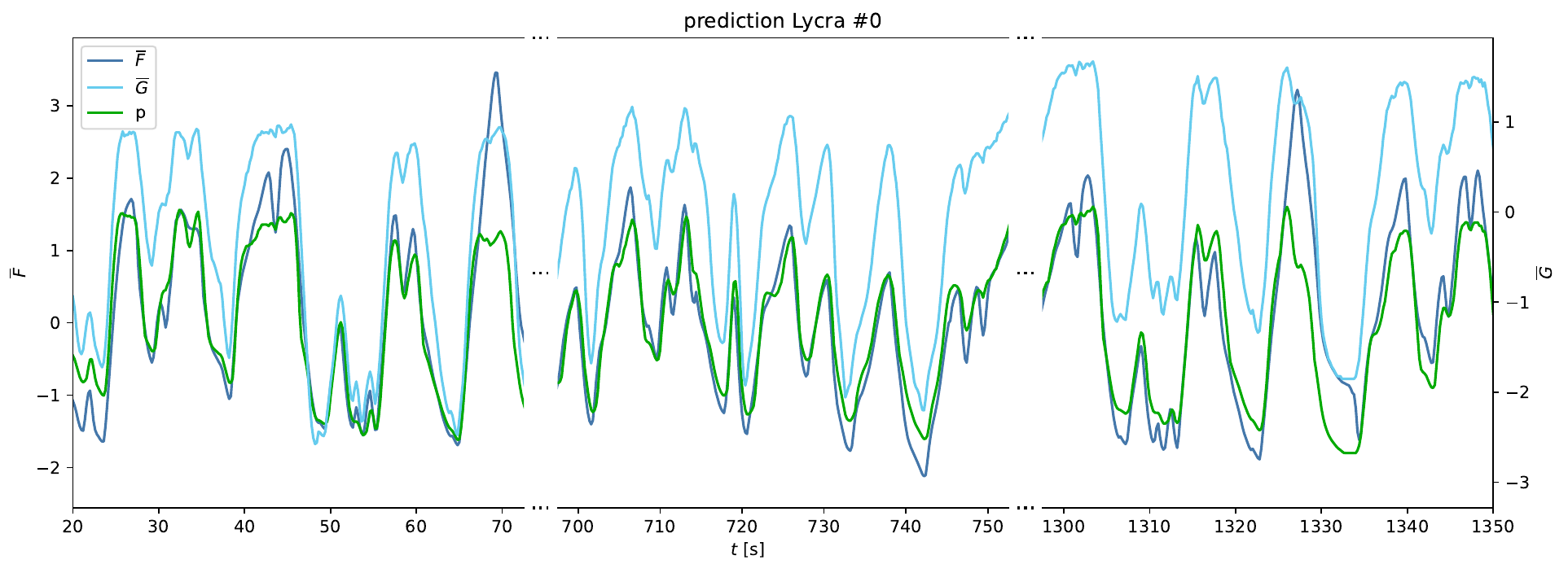}\
        \includegraphics[width=1\columnwidth]{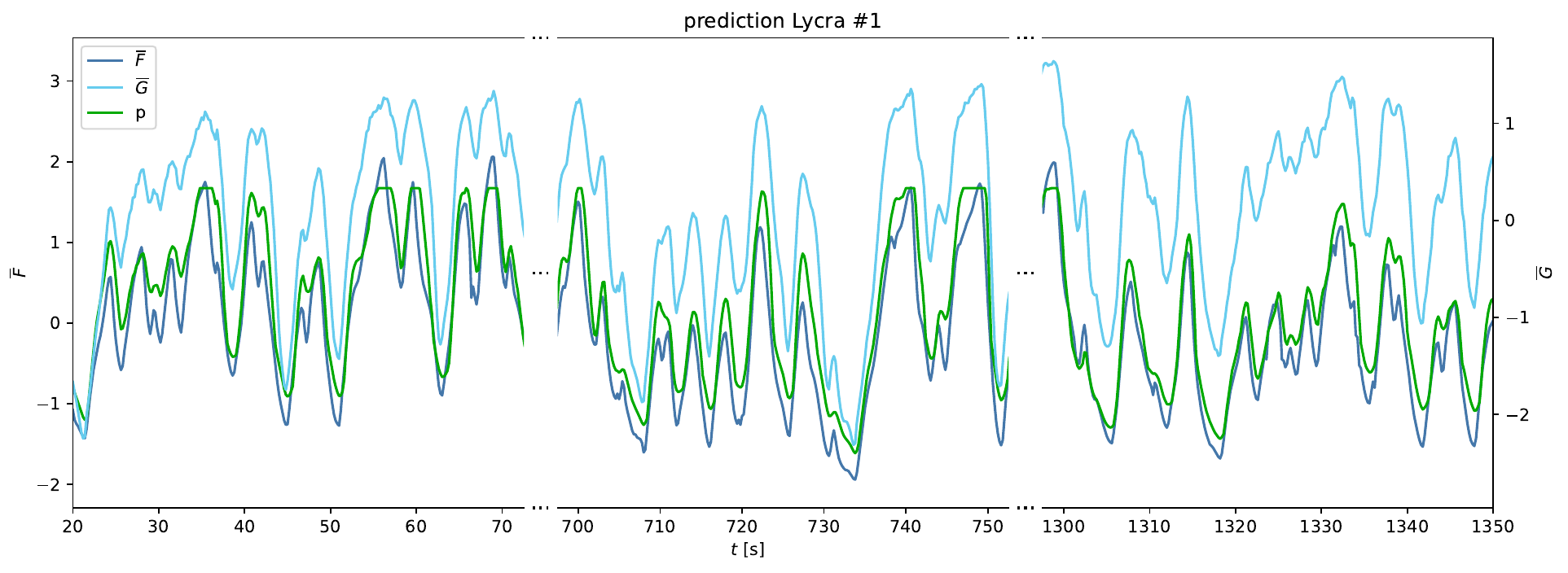}%
    \end{subfigure}
    \caption{Our results show that prediction values $p$ rectify sensor inconsistencies $G$ exceptionally well. Most striking is the removal of long-term drift (top). When looking more closely at segments (left), we see that $p$ aligns well with ground truth data $F$ in most cases. The same method also translates well to our Lycra-variants using different yarn material and knitting composition.}
    \label{fig:prediction}
\end{figure*}

Our systematic tests resulted in the best $r^2$ score for the parameter-combinations of $a$ = 2.5, $N$ = 7 (i.e., $A$ = (0.4, 0.16, 0.064, 0.0256, 0.01024, 0.004096, 0.0016384), HL base size = 4, and topology = (1, \textonehalf, \textonehalf), which gives neuron counts of (4, 2, 2) (cf. Figure \ref{fig:ann}). Predictions of force values from our training sets resulted in $r^2$ values of 0.791 and 0.767 (PES), as well as 0.830 and 0.828 (Lycra), producing highest gains for the PES variants. Figure \ref{fig:prediction} shows the result of test set \#1, with $p$ overlaid on $\overline{G}$ and $\overline{F}$. It is striking that the long-term drift was removed. Inspection of section snippets (cf. Figure \ref{fig:prediction}, left) reveal considerable improvement over the pre-processed input signal that goes well beyond what could be achieved with more basic transformation techniques. Using identical parameters, the network was also trained for the Lycra test-set, and as can be seen in Figure \ref{fig:prediction} (right), there is similar improvement. In general, however, we can observe occasional underestimation of peaks, while the model seems to perform very well at low-force areas. 

\subsection{Further Experiments}

A reasonable question at this point is whether or not our method translates to different kinds of input data. Since the majority of related work utilizes knitted sensors not primarily as \textit{force} sensors but rather as \textit{strain} sensors, we ran our recordings of actuator offset through the same pipeline to see if our technique also translates to this objective. As mentioned above, $d$ also drifts over time which may be the initial cause of this long-term drifting effect, since $G$ does not considerably drift relative to $d$, however both trends still differ significantly, with initial $r^2$-scores between 0.260 and 0.319 (PES), as well as 0.337 and 0.490 (Lycra). Using our unmodified processing pipeline as presented in Figure \ref{fig:ann}, we were able to boost those scores up to 0.699 (PES), and 0.669 (Lycra) for the test sets (further details can be found in the supplement). This implies that our method works exceptionally not only for force as a main metric.

A related question concerns variation of actuation speed and amplitudes, since for our main study, we generated trajectories that feature similar statistic values for (cf. Table \ref{tab:trajectory}), in order to assess consistency of our results. To estimate how well the proposed method translates beyond that specific data, we generated more trajectories with strain values up to 50\%, actuator velocities up to 14\,mm/s (mean: 2.87\,mm/s, SD: 2.06\,mm/s), and accelerations up to 65.54\,mm/s\textsuperscript{2} (mean: 9.27\,mm/s\textsuperscript{2}, SD: 7.87\,mm/s\textsuperscript{2}) and recorded data using the PES patches. We noticed that initial r\textsuperscript{2} scores were already higher (median: 0.644), suggesting that sensors are more consistent for higher actuation speeds. With applying our model, we could still boost the r\textsuperscript{2} by 15\% (median). One extreme case was a rise from 0.349 to 0.795, thus a gain of 128\%. However, we judged that $G_1$, i.e., the NN input feature with least smoothing, already lagged too much from the original. We countered this by reducing smoothing, lowering $a$ from 2.5 to 1.75, which resulted in $\alpha_1=0.57$, etc., which resulted in a better r\textsuperscript{2} gain of 24\% (median). Summarizing, this shows that our presented method does translate to different data, however, fine-tuning smoothing factors against an estimated range can still be beneficial. In yet another experiment, we applied a NN trained with data from the initial trajectories to the data with the newly generated ones. Using the initial values for smoothing factors, we were able to increase r\textsuperscript{2} by 23\% (median), meaning already pre-trained networks can also be reasonably applied across different data sets exhibiting different statistic distributions.




In terms of feature choice would like to note that we tried several variations, e.g., including additional information: first, we experimented with slope values, i.e., first derivatives of each of the smoothed signals $G_i$, as well as a set of smoothed first derivatives of $G$, however we found those did not improve prediction quality significantly as they seem redundant. Second, we briefly experimented with features taken from the frequency domain by calculating windowed FFTs, however the signal turned out to contain little information beyond very low frequencies and we did not go into great lengths to exploring this direction any further. In terms of alternative machine learning methods, we experimented with linear\footnote{\url{https://scikit-learn.org/stable/modules/generated/sklearn.linear_model.LinearRegression.html}}, polynomial\footnote{\url{https://scikit-learn.org/stable/modules/generated/sklearn.preprocessing.PolynomialFeatures.html}} (3\textsuperscript{rd} and 4\textsuperscript{th} order), and random forest regressors\footnote{\url{https://scikit-learn.org/stable/modules/generated/sklearn.ensemble.RandomForestRegressor.html}}, but were not able to produce results of comparable quality.

Furthermore, we briefly tried re-sampling to rates other than 20\,Hz, namely half and twice the frequency. We got slightly worse results with 10\,Hz and similar results with 40\,Hz, so we kept our initial value of 20\,Hz for data pre-processing.

\subsection{Discussion \& Limitations}


The results show that our approach performs remarkably well in improving the mapping from sensor reading to physical actuation. The fact that our processing pipeline is exceptionally small and requires little computation effort, makes it viable for applications in highly limited environments and platforms. We would like to note that due to the nature of our method that is based on features from smoothed sensor readings, as opposed to resorting to a learned hysteresis-model, it is reasonable to assume that our method adapts well to permanent effects, such as chemically and mechanically induced material degradation.

We showed that our method proves successful also when applied to sensors with considerable differences in structure and elastic behavior. Furthermore, the network can be trained to successfully predict not only force but also strain values. Further experiments with actuation of varying strain amplitudes and speeds suggest that our method translates well between different input data, although fine-tuning parameters can be beneficial. We believe that this can be evaded by increasing the number of features $G_i$, with smaller smoothing factor steps in-between. This will increase computational complexity and possibly add redundancy, however, the resulting model could be more versatile. However, we believe it may be possible to infer optimal $a$ and $N$ from a given data set with reasonable effort. We see great value in implementing a fully adaptive system this way and plan to investigate in this direction in future work.

We do acknowledge a few limitations of this work. First, we did not go into great length in investigating entirely different ANN topologies, such as Deep Belief Networks \cite{Bengio2006}, Extreme Learning Machines \cite{Huang2006}, Echo State Networks \cite{Jaeger2004}, etc. We do not expect serious performance gains by changing the topology type, however it will be an interesting direction to explore, and we leave this for future work. 

Second, we do not have data that goes beyond our roughly 23 minutes recordings. We did observe in prior evaluations, that drift decreases in a logarithmic manner, therefore we expect the most challenging effect of drift is to be addressed at the beginning. Furthermore, our method presents a multi-purpose and adaptive way of handling the issue in that it provides a means of modeling long-term drift in the form of highly smoothed sensor signals (cf. $G_6$ and $G_7$ in Figure \ref{fig:timeline}, bottom right) and use this as highly compact and low-complexity features. In terms of signal peaks of high prominence that are sometimes underestimated (cf. Figure \ref{fig:prediction} PES \#0 at second 748) or cropped (PES \#1 at second 740), it is reasonable to believe that further training with data that is more specific to this issue will rectify the model accordingly.

Third, we present a method of predicting scaled data. Mapping the range to meaningful physical values will require some initial calibration step. Note that this calibration would also be required without our pipeline, since, e.g., readings in Ohms or Siemens need to be translated to Newtons either way.

Fourth, we mentioned we re-sampled our data to a constant sample-rate, since our exponential smoothing filters are not considering timing data and are therefore sensitive to varying $\Delta t$. We believe this can be easily overcome by ensuring a more consistent sample rate in the firmware and by taking timing data into account for smoothing. As mentioned, the actual ADC and firmware would be able to sample at a much higher rate (128\,Hz in our case) and the solution comes down to implement a solid down-sampling routine, that outputs at a reasonably constant rate. 

Fifth, for this investigation, we initialized our exponential smoothing filters with the mean of the first $M$ samples to avoid biased starting values (cf. Section \ref{sec:processing}). Applying this in a real-world application would require a short duration for initialization for collecting those samples. However, we did not see this as a major limitation and not as the core of this work. We trust it is not a serious challenge to find alternative initialization methods to set $y(0)$ as there are numerous ways to do so \cite{Nahmias2008}, or to substitute the entire smoothing filter, as we do not believe our method relies on this exact method for smoothing.

No doubt, the very best solution to specific sensors may vary in detail, slight modifications of the network and features (e.g., smoothing factors and number of features) may be beneficial for fine-tuning to the scenario at hand. However, we noticed during our experiments that many of those slight adjustments only result in minor improvements that may not be significant or representative and could be subject to the particular training data.
\section{Conclusion}
\label{sec:conclusion}


We demonstrated a method of utilizing an ANN for correcting inconsistent sampling data read from a knitted resistive force sensor, by pre-filtering the raw input signal and thus providing multiple levels of temporal information as a feature vector to the NN. Once trained, the pipeline can be used as a real-time filter for translating sensor readings to physical actuation data. We demonstrated our method using a MLP NN that in its best-performing configuration requires only three hidden layers and a total of 8 neurons to achieve considerable improvement over the input data, which signifies exceptionally low computational requirements and therefore facilitates applications in a wide variety of scenarios. Although we applied the technique to translate raw sensor data to the trend of physical force/stress, the method can be easily translated to related metrics such as strain, as we briefly demonstrate in the supplement. Furthermore, we successfully applied our technique to a knitted sensor of different structure and behavior, which implies it translates well to slightly different conditions, possibly even to entirely different use cases beyond knitted sensors that suffer from similar issues, such as considerable hysteresis and drift. 

\section*{Acknowledgment}
This research is part of the COMET project TextileUX (No. 865791, which is funded within the framework of COMET -- Competence Centers for Excellent Technologies by BMVIT, BMDW, and the State of Upper Austria. The COMET program is handled by the FFG.

\bibliographystyle{ieeetr}
\bibliography{main}
\end{document}


\newcommand{\iko}[1]{\textcolor{blue}{iko: #1}}
\newcommand{\note}[1]{[\hl{#1}]}

\maketitle

\section{Introduction}

This document represents supplementary material for the paper "Machine Learning Based Compensation for Inconsistencies in Knitted Force Sensors" by Aigner and Stöckl \cite{Aigner2023}. The following chapters present in-depth plots as well as more detail on transferability of the proposed method to predicting strain instead of force, which was only touched briefly in the paper, to avoid oververbosity.

\section{Timeline Plots}

\begin{figure*}
    \centering
    \begin{subfigure}{1\textwidth}
        \centering
        \includegraphics[width=1\textwidth]{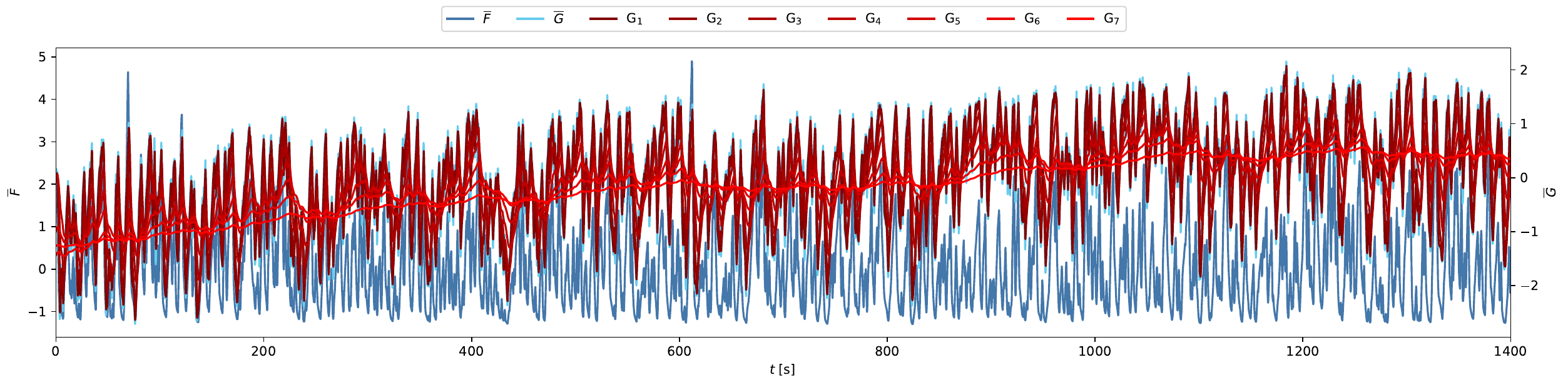}
    \end{subfigure}\
    \begin{subfigure}{1\textwidth}
        \centering
        \includegraphics[width=1\textwidth]{pics/prediction-PES-0-full.pdf}
    \end{subfigure}\
    \begin{subfigure}{1\textwidth}
        \centering
        \includegraphics[width=1\textwidth]{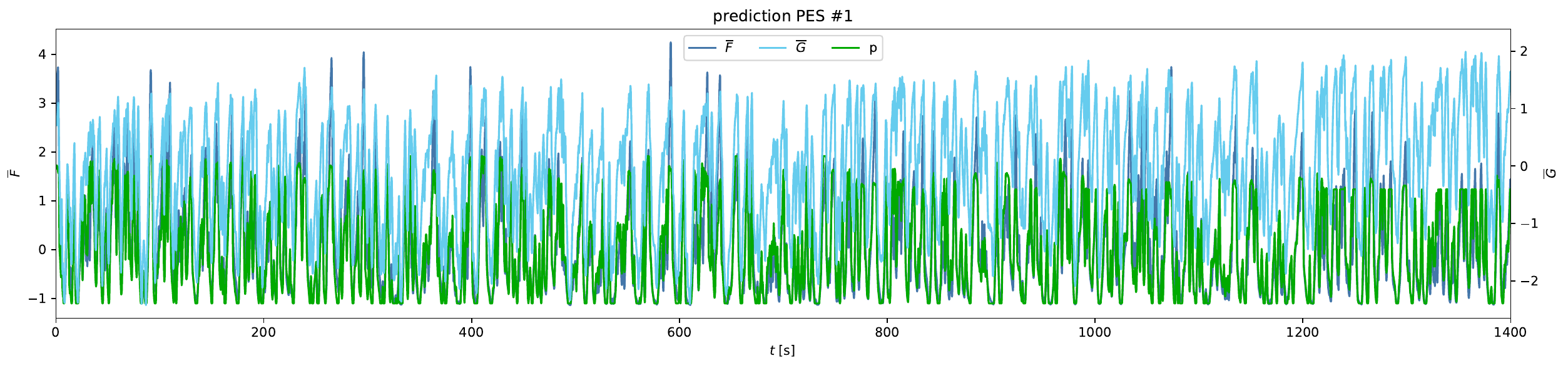}
    \end{subfigure}\
    \begin{subfigure}{1\textwidth}
        \centering
        \includegraphics[width=1\textwidth]{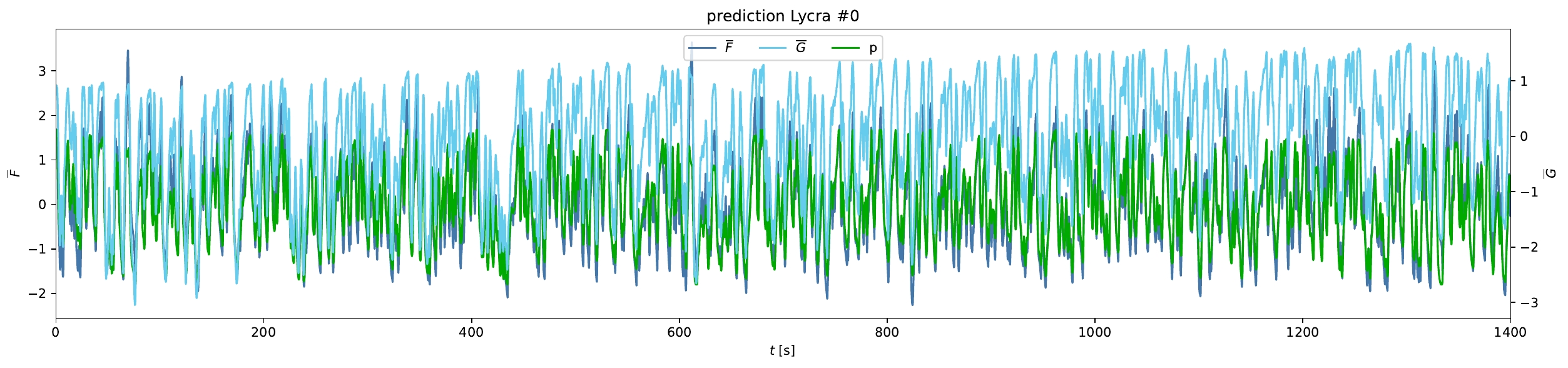}
    \end{subfigure}\
    \begin{subfigure}{1\textwidth}
        \centering
        \includegraphics[width=1\textwidth]{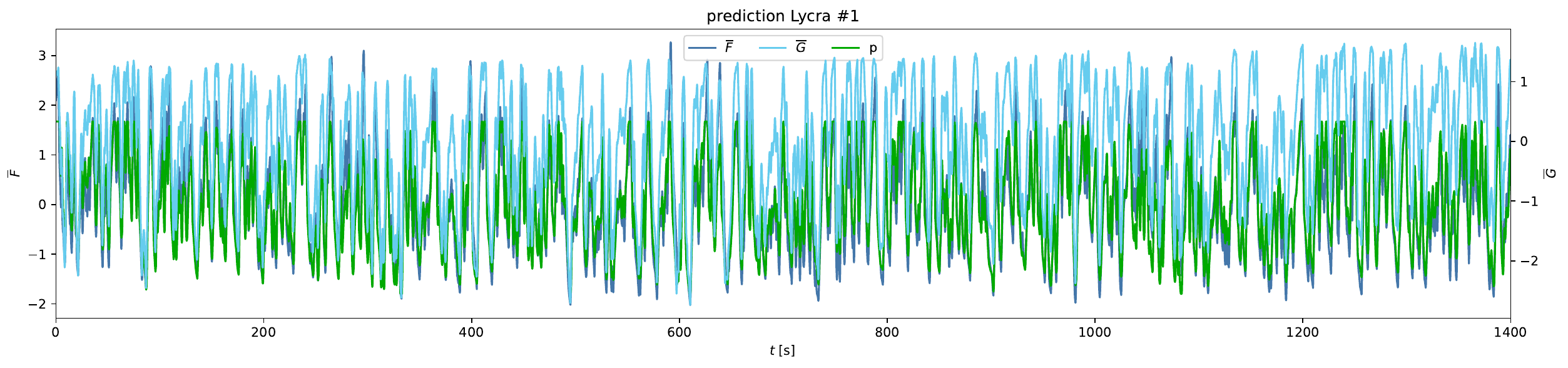}
    \end{subfigure}
    \caption{Features $G_1$ thru $G_7$ of our PES training set (top) and entire prediction results of our test sets.}
    \label{fig:timelines-full}
\end{figure*}

We include additional timeline plots, complementary to the ones in the main paper: Figure \ref{fig:timelines-full} (top) shows input features $G_1$ thru $G_7$ for the entirety of our collected data set. It is clearly visible that features with low $\alpha$ increasingly model the long-term drift. The remaining sub-figures show the full timeline plots of both PES and both Lycra test sets, in addition to the close-ups in the full paper. 

\section{Mapping Sensor Data to Actuator Displacement}

\begin{figure*}
    \centering
    \begin{subfigure}{1\textwidth}
        \centering
        \includegraphics[width=1\textwidth]{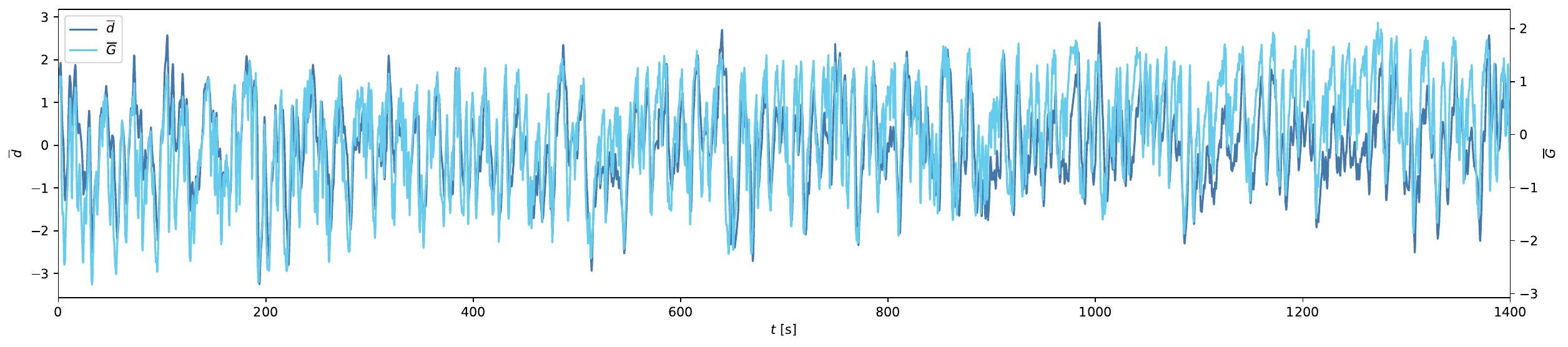}
        \caption{}
    \end{subfigure}\
    \begin{subfigure}{1\textwidth}
        \centering
        \includegraphics[width=1\textwidth]{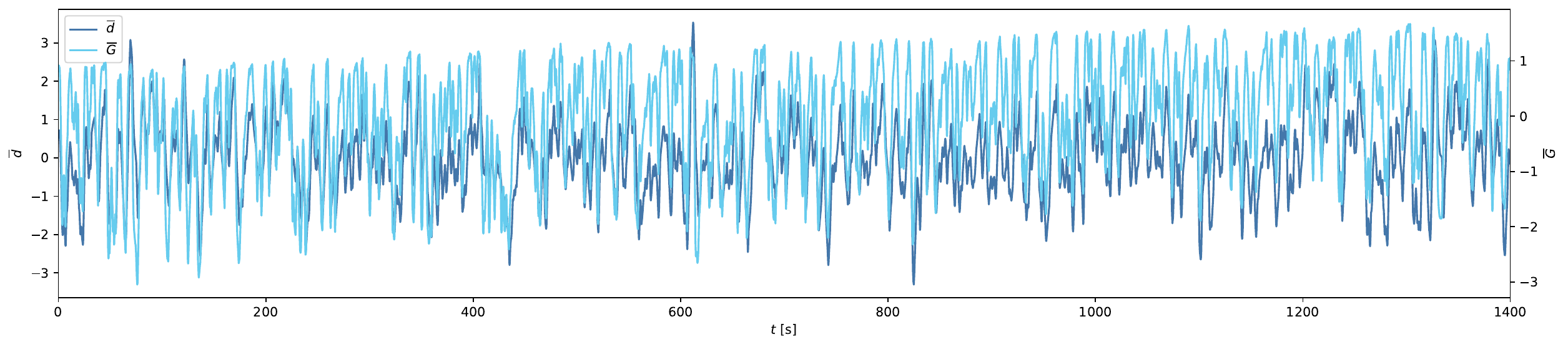}
        \caption{}
    \end{subfigure}\
    \begin{subfigure}{0.5\textwidth}
        \centering
        \includegraphics[width=1\textwidth]{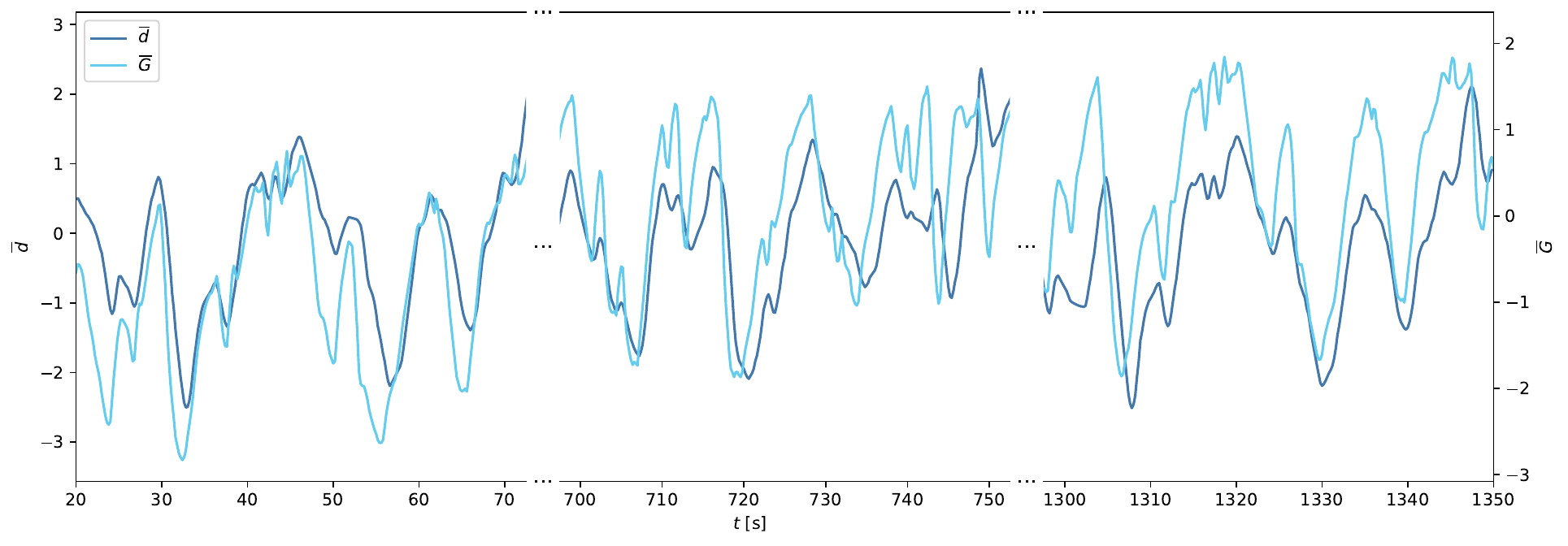}
        \caption{}
    \end{subfigure}%
    \begin{subfigure}{0.5\textwidth}
        \centering
        \includegraphics[width=1\textwidth]{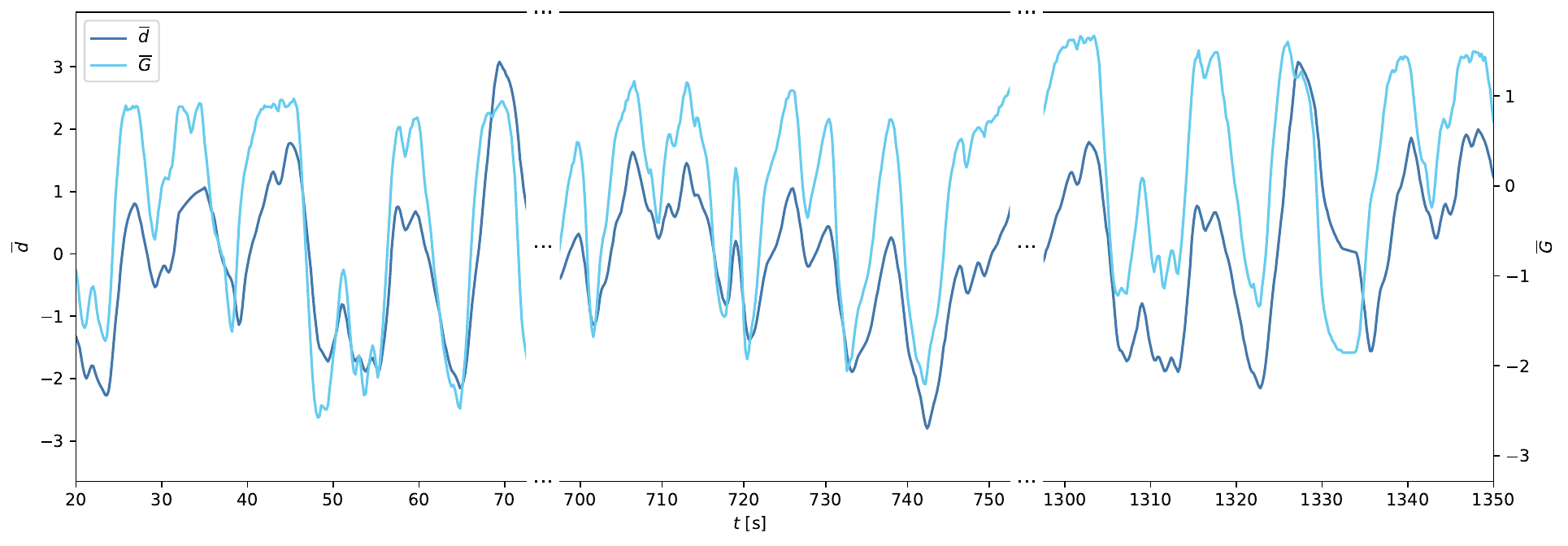}
        \caption{}
    \end{subfigure}
    \caption{Plots showing $\overline{d}$ and $\overline{G}$ of PES \#0 (a,b) and a Lycra \#0 (c,d) test sets.}
    \label{fig:timelines-d}
\end{figure*}

As briefly touched in the paper, our technique works for strain/displacement data as well. To achieve the following results we merely trained against normalized displacement $\overline{d}$, by re-sampling actuator displacement (i.e., absolute sensor elongation) $d$ to 20\,Hz and then removing mean and scaling to unit variance using the \textit{StandardScaler} from the \textit{scikit learn} Python package\footnote{\url{https://scikit-learn.org/stable/modules/generated/sklearn.preprocessing.StandardScaler.html}}. We used the exact same pipeline including initialization of $y(0)$ for exponential smoothing filters, as well as neural network hidden layer design, MLPRegressor activation function, etc. Figure \ref{fig:timelines-d} shows timeline plots of an exemplary PES (top) and Lycra recordings (bottom). 

We can see that $d$ seems to drift along with $G$, however the relative drift differs in between sensor variations. Our method adapts well to these differences: $r^2$ values (pre- and post-prediction) can be found in Table \ref{tab:prediction}, which shows considerable gain in mapping between sensor conductivity and elongation when applying our method, with highest gains for PES patches. 

Note that the model was not at all manually adapted to the different objective; in preliminary experiments, we found that by changing the NN's hidden layers, we could slightly improve test scores up to 0.716. However, we believe those minor differences are subject to the training set and do not make a crucial difference in real-world applications.






\begin{figure*}
    \centering
    \begin{subfigure}{0.5\textwidth}
        \centering
        \includegraphics[width=1\textwidth]{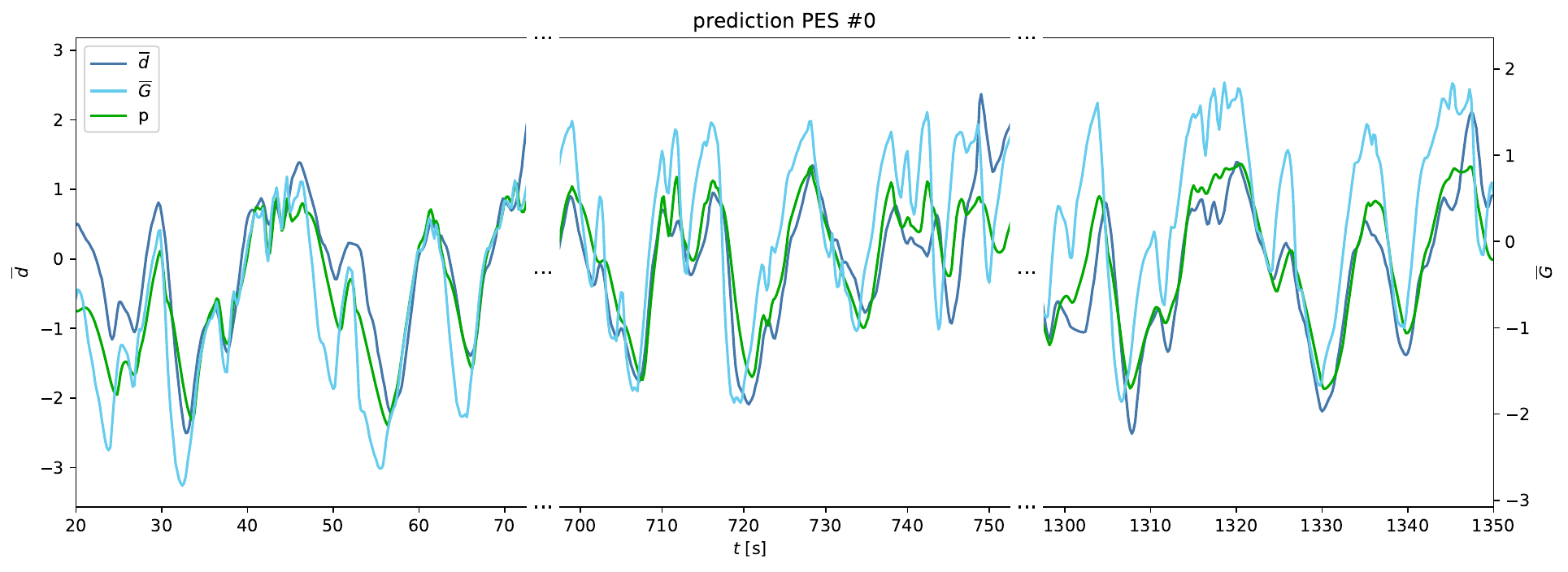}
    \end{subfigure}%
    \begin{subfigure}{0.5\textwidth}
        \centering
        \includegraphics[width=1\textwidth]{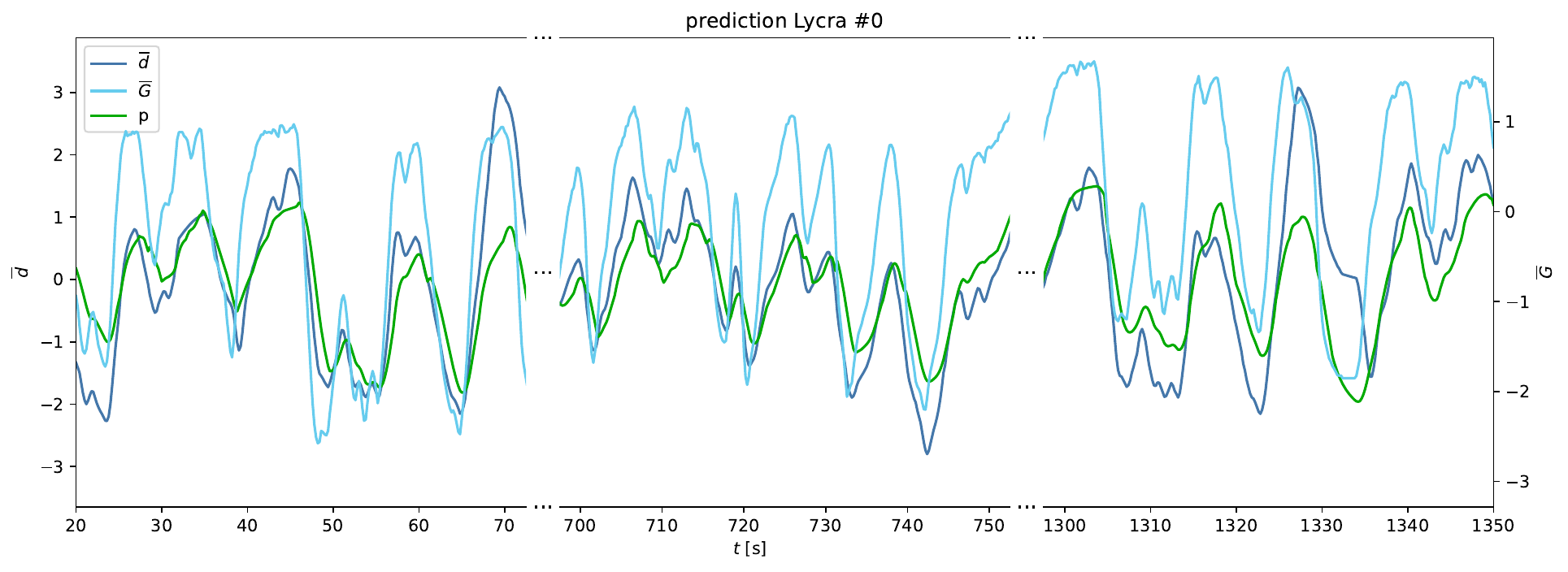}
    \end{subfigure}\
    \begin{subfigure}{0.5\textwidth}
        \centering
        \includegraphics[width=1\textwidth]{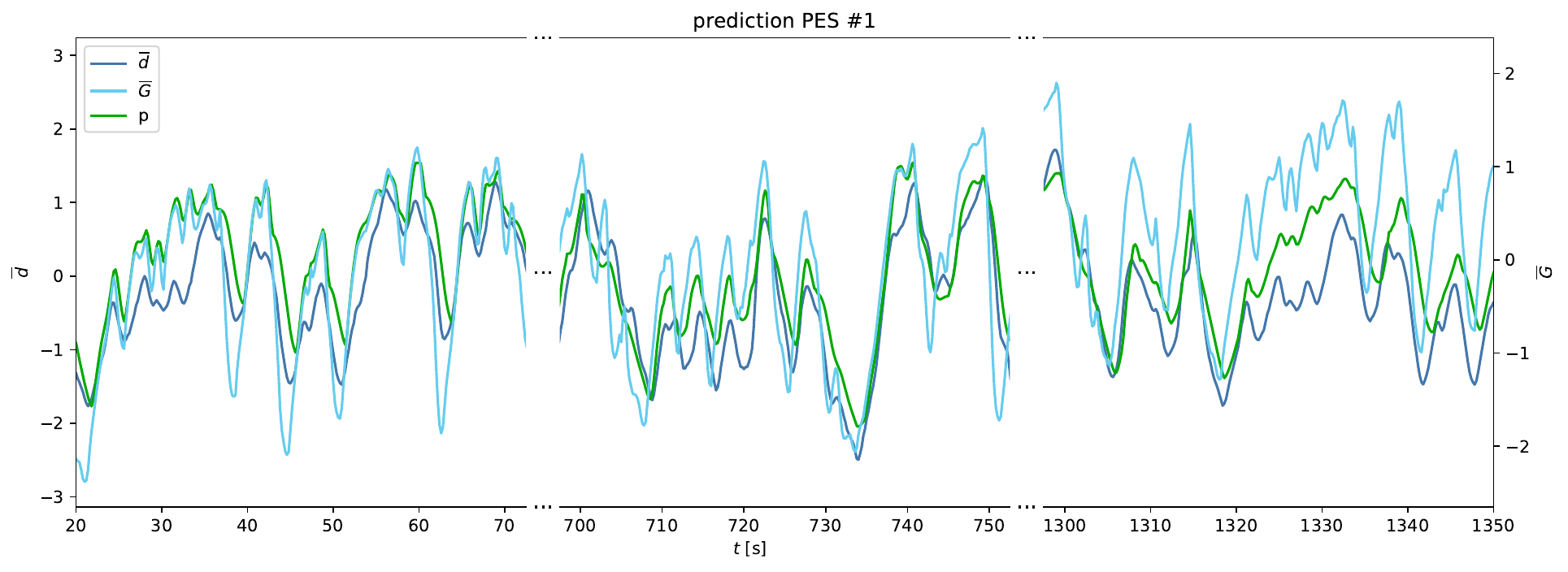}
    \end{subfigure}%
    \begin{subfigure}{0.5\textwidth}
        \centering
        \includegraphics[width=1\textwidth]{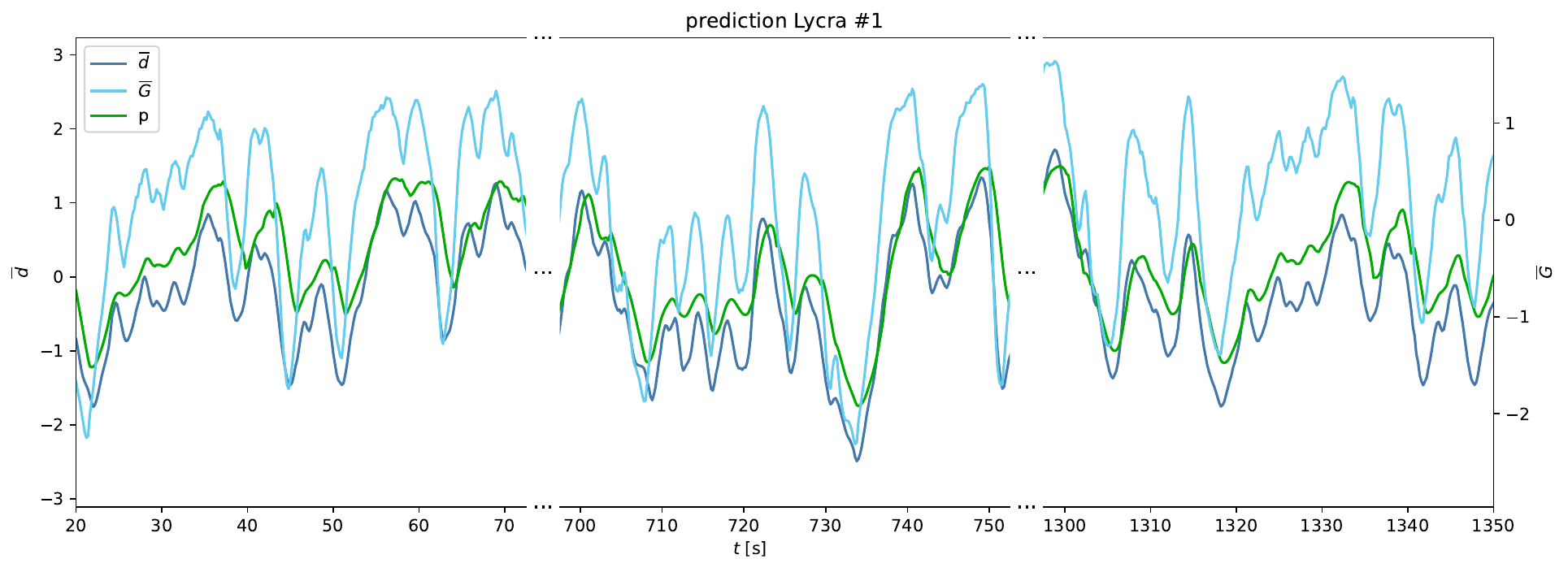}
    \end{subfigure}
    \caption{Resulting predictions $p$ show good rectification of the input signals for all of our test sets, although we can observe occasional under-estimations of peak areas in the signal, as is the case when predicting force data (see main paper).}
    \label{fig:timelines-prediction-d}
\end{figure*}

\section{Notes on Enclosed Spreadsheet}

In order to compare model in terms of their performance on \textit{both} test sets A and B, we calculated an error metric $E$ from the respective $r^2$ scores with 
    $$E = \frac{(1-r^2_A)^2 + (1-r^2_B)^2}{2} \,.$$
Values reported and color-coded in the enclosed spreadsheet \texttt{nn-eval.xlsx} represent according $E$-values, i.e., values close to 0 indicate better performance.

\clearpage

\begin{table}
    \centering
    \caption{$r^2$ of our initial (pre-processed) and predicted data when applied to actuator displacement. We included results of our two test sets (A, B), as well as the training sets (T) for sake of completeness. PES show highest gain from our approach, however, the Lycra patches ultimately yield higher scores.}
    \begin{tabular}{L{0.1\columnwidth}|C{0.15\columnwidth}|C{0.15\columnwidth}|C{0.15\columnwidth}}
         & r²(d,G) & r²(d,p) & gain \\
        \hline \hline
        PES\textsubscript{t}   & 0.319 & (0.705) & (0.386) \\
        PES\textsubscript{A}   & 0.319 & 0.699 &  0.380  \\
        PES\textsubscript{B}   & 0.260 & 0.635 &  0.375  \\
        \hline
        Lycra\textsubscript{t} & 0.337 & (0.687) & (0.350) \\
        Lycra\textsubscript{A} & 0.479 & 0.609 &  0.130  \\
        Lycra\textsubscript{B} & 0.490 & 0.669 &  0.179
    \end{tabular}
    \label{tab:prediction}
\end{table}

\section{Data Processing Pipeline}

\begin{figure}[h!]
    \centering
    \includegraphics[width=1\textwidth]{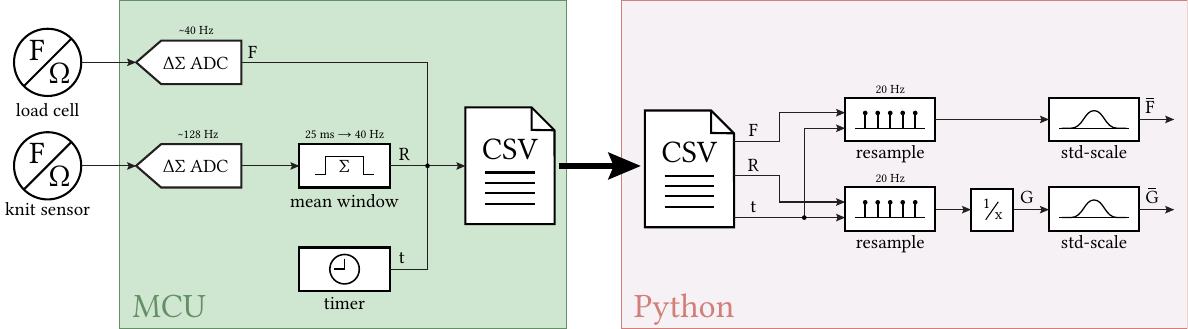}
    \caption{During data recording, values of knitted sensors are super-sampled and in order to reduce measurement noise. Within time windows of 25\,ms, mean values are calculated and written to CSV files, along timestamps and data sampled from the load cell. During pre-processing for training our ML models, data is resampled to unified time periods between samples, since our proof-of-concept does not yet take timing into account, for reasons of simplicity.}
    \label{fig:data-processing}
\end{figure}

Figure \ref{fig:data-processing} shows the data processing steps that are involved for both acquiring data from the sensors on the MCU, as well as during pre-processing for training the ANNs in Python. Load cell as well as knitted sensors were sampled using two Delta-Sigma ADCs by a single ESP32 MCU. Resistance values read from the load cell were converted to Newtons already in the firmware. Since resistance readings of the textile sensor were slightly noisier, we supersampled with 128\,Hz, buffered values and calculated mean values in the firmware every 25\,ms. Since timing on the firmware could not be controlled to achieve periods of 25\,ms precisely, we resampled to constant frequency in Python later based on the timestamps that were recorded to the CSV file along the sensor readings. Furthermore, resistance values $R$ were inverted to get conductivity values $G$. To achieve better performace of the machine learning estimators to be used, we normalized both $F$ and $G$ uniform ranges using the scikit learn StandardScaler\footnote{\url{https://scikit-learn.org/stable/modules/generated/sklearn.preprocessing.StandardScaler.html}}, which centers data round $\mu$ and scales with $1/\sigma$.

\bibliographystyle{IEEEtran}
\bibliography{IEEEabrv,supplement}